\documentclass[12pt]{article}
\usepackage[pdftex]{graphicx}
\usepackage{color}
\usepackage{fullpage}
\usepackage{amsmath,amsthm,amssymb}
\newcommand{\T}{^{\mathsf{T}}}
\newcommand{\rmd}{\mathrm{d}}
\newcommand{\rme}{\mathrm{e}}
\newcommand{\arcl}{s}

\newcommand{\pth}{Z}
\title{Maximum Flux Transition Paths of Conformational Change}

\author{Ruijun Zhao, Juanfang Shen, and Robert D. Skeel \\
Department of Computer Science, Purdue University \\
West Lafayette, IN 47907-2107}
\date{December 22, 2009}

\begin{document}\maketitle


\begin{abstract}
Given two metastable states $A$ and $B$ of a biomolecular system,
the problem is to calculate the likely paths of the transition from $A$ to $B$.
Such a calculation is
more informative and more manageable if done for a reduced set of collective
variables chosen so that paths cluster in collective variable space.
The computational task becomes that of computing the ``center'' of such a
cluster.
A good way to define the center employs the concept of a committor, whose value
at a point in collective variable space is the probability that a trajectory at
that point will reach $B$ before $A$.
The committor ``foliates'' the transition region into a set of isocommittors.
The maximum flux transition path
is defined as a path that crosses each isocommittor at a point which
(locally) has the highest crossing rate of distinct reactive trajectories.
(This path is different from that of the MaxFlux method of Huo and Straub.)
It is argued that such a path is nearer to an ideal path
than others that have been proposed with the possible exception
of the finite-temperature string method path.
To make the calculation tractable, three approximations are introduced,
yielding a path that is the solution
of a nonsingular two-point boundary-value problem.
For such a problem, one can construct a simple and robust algorithm.
One such algorithm and its performance is discussed.
\end{abstract}


\section{Summary}

Considered here is the problem of computing transition paths
of conformational change, given two different metastable states of a
biomolecule.
One motivation for this is to facilitate the accurate calculation of
free energy differences.
Another motivation is to determine the existence and structure of
transition states and intermediate metastable states.
The latter are possible targets for inhibitors of enhanced specificity
in cases where a family of proteins have active sites with
very similar structure.
A good example of this situation is the Src tyrosine kinase family~\cite{ZhYG09},
which has long been implicated in the development of cancer.
For this system there are already computational
results~\cite{GaYR09,OzPo06,YaBR09}, supported by experiment~\cite{OYMP08},
for the transition path from an active catalytic domain to an inactive catalytic domain.

Some approaches to this problem generate ensembles of
trajectories based on the equations of motion.
Notable examples are transition path sampling~\cite{BCDG02} and Markov state models~\cite{SiPa05}.
Applying such methods to large proteins (without compromise)
would appear to require exceptional computing capabilities,
so here we pursue a more theoretical approach
that avoids ``direct numerical simulation.''
Such approaches, like those in Refs.~\cite{HuSt97,OlEl96,PSLS03},
seek to characterize one (or several isolated) ``representative''
reaction paths connecting two given metastable states,
each path representing a bundle or cluster of trajectories.
Here we adopt a well developed and tested theory,
namely, transition path theory (TPT)~\cite{EVa06,MFVC06,MeSV06,Vand06}.
Additional references on computing transition paths are found in Ref.~\cite{PaSR08}.
In general, it may also be
of interest to calculate (i) the reaction rate for each bundle,
or, at least, the relative rate for different bundles,
and (ii) the potential of mean force.
Here we consider only the calculation of the path itself.

In a nutshell, this article embraces a certain aspect of TPT
and carries it to a logical conclusion,
obtaining a formula, an implementation, and a proof of concept.
The claim is that we can compute a path that is closer to the ideal
than the minimum free energy path (MFEP)~\cite{MFVC06}
and that, in a couple of respects,
is better than the path of the finite-temperature
string (FTS) method~\cite{RVME05, VaVe09}, though inferior in another (important) respect.
Additionally, the formula for the path is computationally more attractive
than the formula that underlies
either the path of the FTS method {\em or the MFEP}.

\subsection{Outline and discussion}

There are two distinct steps in getting a solution:
The first is to define the problem without concern for the methods to
be employed (other than taking into account the intrinsic difficulty of the problem).
Defining a problem apart from a method gives a more concise definition.
Also, by not guessing about what is feasible computationally,
one may avoid unnecessary compromises.
The second step is to construct a method and algorithm.

Given two metastable states $A$ and $B$ of a biomolecular system,
the aim is to calculate the likely paths of the transition from $A$ to $B$.
Such a calculation is more informative and more manageable
if done for a reduced set of {\em collective variables},
functions of the system configuration $x$,
\[ \zeta_1 =\xi_1(x), \zeta_2 = \xi_2(x), \ldots, \zeta_\nu =\xi_\nu(x),
 \quad\mbox{abbreviated as }\zeta = \xi(x),\]
chosen so that paths cluster in collective variable space.
The computational task
becomes that of computing the ``center'' of such a cluster.
A good way to define the center employs the concept of a committor,
whose value at a point in collective variable space
is the probability that a trajectory at that point will reach $B$ before $A$.
The committor ``foliates'' the transition region into a set of
committor isosurfaces known as isocommittors.
The maximum flux transition path (MFTP)
is defined as a path that intersects each isocommittor at a point which
(locally) has the highest crossing rate of distinct reactive trajectories.
The MFTP is not to be confused with MaxFlux method~\cite{BMMN83, HuSt97};
it differs in several respects, in particular, the MFTP considers
the flux of only those trajectories that are reactive
(by using a result from TPT).
A more detailed account of the problem definition
is given in Section~\ref{sec:path}.

The minimum free energy path has been used for some time
to represent reactive trajectories in collective variable space.
Only fairly recently
has its relationship to reactive trajectories been explained.
The article~\cite{MFVC06} applies large deviation theory
to show that the MFEP is the most probable path in the zero
temperature limit of dynamics on a free energy surface defined
at finite temperature.
Hence, the MFEP is an inherently inconsistent construct and it is {\em useful only to the extent
 that it represents fully finite-temperature trajectories}.
In fact, it does this fairly well on the simple tests reported here.
Other fully finite-temperature constructs have been proposed:
the finite-temperature string method
in collective variable space (Sec.~IV.B of Ref.~\cite{RVME05})
and the swarm-of-trajectories string method~\cite{PaSR08},
which constructs a Brownian dynamics model on the fly
and constructs a path whose tangent is the most probable direction.
How they differ from the MFTP is detailed in Section~\ref{sec:path}.

To make the calculation tractable,
three approximations are introduced.
To make the committor a more accessible quantity,
the set of paths is approximated by a Brownian dynamics model,
resulting in a boundary value problem in $\nu$-dimensional space.
Then the number of space dimensions is reduced to one by assuming
most of the transition paths are contained in a tube,
resulting in a two-point boundary-value problem with $2\nu$ unknowns.
A third approximation reduces this to $\nu$ unknowns,
whose solution is a {\em maximum flux transition path}.
The resulting equations involve a free energy gradient term
and an explicitly temperature-dependent curvature term.
Specifically, the maximum flux transition path $\zeta = Z(s)$, $0\leq s\leq 1$,
is defined by the condition that
\[
 - \beta\nabla F(Z)
 + \frac{(D(Z)^{-1}Z_s)_s}{Z_s\T D(Z)^{-1}Z_s} ~\parallel~D(Z)^{-1}Z_s,
\]
holds for $\zeta = Z(s)$ where $\beta$ is the inverse temperature,
$F(\zeta)$ is the free energy profile, $D(\zeta)$ is a proto-diffusion tensor
depending on masses and $\xi$, and the subscript $s$ denotes
differentiation $(\rmd/\rmd s)$.
In the high temperature limit, the path becomes a straight line.
In the low temperature limit, the path becomes
an MFEP.
At zero temperature the path will have cusps at some intermediate
local minima, which presents difficulties if free energy profiles
or relative reaction rates are to be determined.
This formula is a key result of this article.
Details are given in Section~\ref{sec:meth}.

The temperature-dependent curvature term not only provides
a finite temperature correction to the MFEP,
but it yields a nonsingular second order ordinary differential equation,
amenable to standard techniques---except for the need to do
computationally intensive sampling to evaluate terms in the differential
equation.
An existing set of algorithms for the MFEP~\cite{ERV07,MFVC06} applies
equally well to the MFTP.
The equation is discretized using upwinded differencing and solved using
the semi-implicit simplified string method~\cite{VaHe08}.
(A notable alternative is the nudged elastic band method,
introduced in Ref.~\cite{JoMJ98}.)
Algorithmic details are provided in Section~\ref{sec:alg}.

Section~\ref{sec:test} compares the MFTP to the MFEP on numerical examples.
First, an artificial problem in full configuration space is solved
to demonstrate the effect of the curvature term of the MFTP.
(A problem in full configuration space is equivalent to a problem
in collective variable space with perfect sampling.)
In particular, the necessity of using an adaptive mesh for the
MFEP is demonstrated.
Then alanine dipeptide in vacuum is solved using the $\phi$, $\psi$ dihedrals
as collective variables.
For the transition path from $C_{7\mathrm{ax}}$ to $C_{7\mathrm{eq}}$ as
in Ref.~\cite{MFVC06}, the computational cost for calculating
the MFTP and the MFEP is almost same.
However, for a transition path from $C_{7\mathrm{eq}}$ to $C'_{7\mathrm{eq}}$
through $C_{7\mathrm{ax}}$ shown in Ref.~\cite{RVME05},
the MFEP has a cusp at $C_{7\mathrm{ax}}$ and the computational cost for
finding such a cusp is expensive.
On the other hand, the MFTP smooths out the cusp
and the computational cost is reduced.

An open source implementation of the MFTP method is available~\cite{Zhao09}
as a relatively simple set of Python modules with examples
using pure Python, CHARMM~\cite{BBMN09}, and NAMD~\cite{NAMD}.

\subsection{Conclusions}

For alanine dipeptide, the MFEP, MFTP, and FTS method paths are quite similar.
On a contrived problem with a rough energy landscape,
e.g., Figure~2 in Ref.~\cite{VaVe09}, the FTS method path gives a much better result.
On a different contrived problem given in Section~\ref{sec:arti},
the MFTP gives a much better result.
Contrived examples are relevant because
computational techniques are sometimes applied in extreme situations
for which they may not have been designed.
In terms of quality, the MFTP ranks higher than the MFEP
but lower than the FTS method path (because the latter addresses
the more serious difficulty of multiple local minima).

The minimum free energy path (and that of the FTS method)
can have cusps at some intermediate metastable states,
which makes it unsuitable for defining an isocommittor,
unsuitable for defining a reaction coordinate, and harder to compute.
Computational difficulties include the need for an adaptive mesh
and a greater number of iterations until convergence.


\section{What is the problem?}  \label{sec:path}

We begin by defining an ensemble of transition paths from $A$ to $B$:
For simplicity, assume the molecular system obeys Newtonian dynamics
with potential energy function $U(x)$ and
a diagonal matrix $M$ of atomic masses.
Positions $x$ and momenta $p$ satisfy $x = X(t)$, $p = P(t)$ where
$(\rmd/\rmd t)X(t) = M^{-1}P(t)$ and $(\rmd/\rmd t)P(t) = -\nabla U(X(t))$.
Initial values are drawn from a Boltzmann-Gibbs distribution $\rho(x, p)$:
positions $x$ from
probability density $\mathit{const}\,\rme^{-\beta U(x)}$
and momenta $p$ from a Maxwell distribution.
Imagine an extremely long trajectory.
The trajectory enters and leaves $A$ and $B$ many times
yielding a huge set of reactive paths from $A$ to $B$.
(A reactive path is a piece of the trajectory outside of $A$ and $B$
that comes from $A$  and goes to $B$.)

Generating an ensemble of trajectories is extremely demanding computationally.
And, even if this were possible, what would the user do with all the data?
By answering such a question,
we might well avoid the task of computing trajectories.
It is likely that one would cluster the trajectories
to produce a concise description.
Therefore, one might instead directly determine such a concise description.
Specifically, if the paths cluster
into one or several distinct isolated bundles/tubes/channels/pathways,
one might compute  a ``representative path'' for each cluster.
This idea is developed in the paragraphs that follow.

However, transition paths
might not cluster adequately---in full configuration space.
Assume, though, there is a smaller set of collective variables,
$\zeta = \xi(x)$, such that in $\zeta$-space,
paths cluster into one or several distinct isolated channels
connecting two separated subsets $A_\xi$ and $B_\xi$
of collective variable space.
Otherwise, there is little of interest to compute.
A typical example of collective variables
is $\phi/\psi$ angles along a peptide backbone.
Once the collective variables are specified, the problem is to
calculate a path in collective variable space,
$\zeta = \pth(\arcl)$, $0\leq\arcl\leq 1$,
connecting $A_\xi$ to $B_\xi$ where the transition paths are concentrated.
Along with a parameterization of the path in collective variable coordinates,
would be a realization of it in cartesian coordinates,
so once the path is generated, structures can be studied as well.
A drawback of this approach is the need
to identify an appropriate set of collective variables.
Indeed, defining suitable collective variables
is an important research problem~\cite{MaDi05}.

We want a minimal set of collective variables subject to
two conditions:
First, the coordinates $\zeta$ must suffice to
describe states $A_\xi$, $B_\xi$
in $\zeta$-space corresponding to $A$, $B$.
Second, coordinates $\zeta$ must also be rich enough to ``express
the mechanism of conformational change'' along the transition path.
To make the second condition more precise, we introduce the notion of
``quasi-committor.''

To measure the progress of a transition,
there is a natural reaction coordinate, known as the {\em committor\/}.
This concept of a commitment probability was introduced
by Onsager~\cite{Onsa38}, and
the abbreviated term ``committor'' was introduced in Ref.~\cite{BoDC00}~(p.~9236),
which they defined as follows:
For each point $x$ in configuration space,
consider a trajectory starting with $X(0) = x$
and velocities drawn at random from a Maxwell distribution,
and define the committor $q(x)$
to be the probability of reaching $B$ before $A$.
Since it is the coordinates of the collective variables that are of
interest, it is natural also to define a {\em quasi-committor\/}:
For each point $\zeta$, consider a trajectory starting with random
initial values {\em conditioned on $\xi(x) = \zeta$}
 and define the quasi-committor $\hat{q}(\zeta)$
to be the probability of reaching $B_\xi$ before $A_\xi$:
\[ \hat{q}(\zeta) =
\Pr(\xi(X(t)) \mbox{ reaches }B_\xi\mbox{ before }A_\xi~|~\xi(X(0)) = \zeta).
\]

\begin{figure}[htbp]
\begin{center}
\includegraphics[width=3.375in]{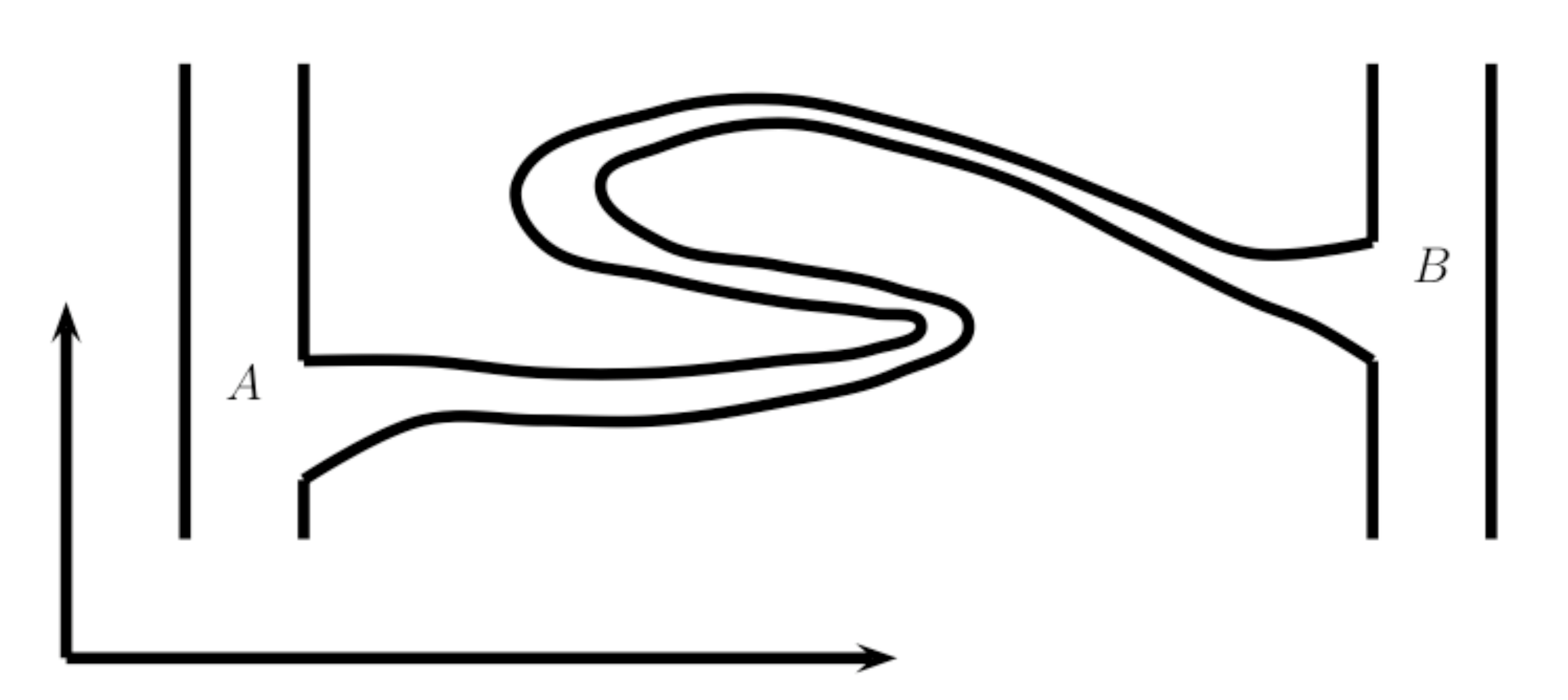}
\caption{ A schematic illustration of a poor choice of collective variables.
The horizontal axis is collective variables, and the vertical axis is
unrepresented degrees of freedom.
The collective variables fail to indicate the progress of the reaction. }
\label{fig:cv}
\end{center}
\end{figure}

We could say that the variables $\zeta = \xi(x)$ are rich enough to
express the mechanism of conformational change if
the quasi-committor $\hat{q}(\zeta)$ has no local minima or maxima
outside of $A_\xi$ and $B_\xi$
(except for regions of negligible probability).
Otherwise, there is some unexpressed degree of freedom
important to the transition.
As an example, suppose that virtually all trajectories stay
within a narrow tube having a geometry in full configuration
space illustrated by \ref{fig:cv}.  Suppose that the free energy profile
{\em as a function of arc length} along the transition tube
is much higher in the backward section than it is in the two forward sections.
Then most of the increase in the quasi-committor as a function of arc length
occurs in the middle section of the tube.
Consequently, the variation in the quasi-committor,
as a function of the ill-chosen
collective variable $\zeta$ corresponding to the horizontal axis,
will be dominated by this middle section of the tube.
This results in a graph of $\hat{q}(\zeta)$ that increases at
the beginning and end of its range but decreases in the middle part.
In addition to $\hat{q}(\zeta)$ having no local extrema, it is desirable that $\hat{q}(\xi(x))\approx q(x)$.
The quality of the collective variables can be checked in principle
by calculating quasi-committor values $\hat{q}(\zeta)$ at points along the path
from dynamics trajectories.

Two approaches have been proposed
for defining the center of a cluster of paths in $\zeta$-space:
\begin{itemize}
\item[(i)] a most probable path,
e.g., a swarm-of-trajectories string method~\cite{PaSR08,Vand08}
path, and
\item[(ii)] a path that intersects each isosurface of the quasi-committor
at a center of the collection of points
where reactive trajectories cross that isosurface,
e.g., a finite temperature string method~\cite{RVME05} path
and a {\em maximum flux transition path}.
\end{itemize}
An MFEP is a limiting case of both approaches.
(The MFEP is obtained from these various approaches by letting
$\beta\rightarrow\infty$ in the path formula but not in the definition
of the free energy profile.)
Defining the objective is a compromise between
(i) best capturing the object of interest and (ii) simplicity.

One problem with seeking the most probable path
is that it is unclear how to assign relative probabilities to paths.
More importantly,
the most probable path tends to be a path of minimum energy,
and it is not clear---a priori---that this is a ``representative'' path.
For Hamiltonian dynamics, it would seem that the
probability that we attach to a path would be proportional to $\exp(-\beta E)$
where $E$ is the energy.
Hence, the most probable path is the one with just
enough energy to surmount the potential energy barriers.
For stochastic dynamics, the explanation of how to assign
probability to paths is quite complicated---if paths
of different durations are being compared.
An explanation for Brownian dynamics is possible using Freidlin-Wentzell theory
and the assumption of vanishingly small noise (see Appendix~A of Ref.~\cite{MFVC06}).
It is reassuring though that the results of Freidlin-Wentzell theory
agree with those of TPT in the zero-temperature limit (for $F(\zeta)$ held fixed).

For defining a path in terms of an intersecting point on each isosurface
of a quasi-committor, one needs
\begin{itemize}
    \item[(i)] a definition for the distribution of crossing points
of reactive trajectories through a quasi-commitor isosurface and
    \item[(ii)] a definition of centrality, e.g., mode, median, or mean.
\end{itemize}
We consider each of these in turn.

The finite-temperature string method defines
the distribution of crossing points of reactive trajectories
in a way that includes recrossings.
A subsequent article~\cite{MeSV06}
illustrates the dramatic distortions that arise by including recrossings,
and it emphasizes crossings of a surface by distinct reactive trajectories
instead of all crossings by reactive trajectories.
They define such a distribution in terms of the net crossings
of reactive trajectories across each infinitesimal piece of a surface.
It is not obvious, however, that this necessarily
gives nonnegative values on the isosurface of a quasi-committor,
so, instead, we use the density of last crossings by reactive trajectories,
called {\em last hitting points} in Ref.~\cite{EVa06}.
For the Brownian dynamics approximation developed in the next section,
these two measures are identical.

Consider now the question of defining the center.
Let $j(\zeta)$ denote the density
associated with a definition for the distribution of crossing points
of reactive trajectories through a quasi-committor isosurface.
One choice for the center is the point of highest probability,
In other words, seek the path
$\zeta = Z(s)$, $0\leq s\leq 1$,
each of whose points $Z(s)$ is a local maximum of the density $j(\zeta)$
on the quasi-committor isosurface $\Sigma$ passing through $Z(s)$.
This is what we use for the MFTP.
Another choice, associated with the {\em finite-temperature
string method}, is to construct the path from the mean value $\zeta'$
on each quasi-committor isosurface $\Sigma$:
the point $\zeta'$ that minimizes
$\int_\Sigma |\zeta' - \zeta|^2 j(\zeta)\rmd\zeta$.
Although this notion is a superior measure of centrality,
it is more complicated to explain.
In practice, methods for finding a maximum are designed only
to find a local maximum, which is what we do for the MFTP.
This is satisfactory if there is a choice of collective variables
that produces a free energy landscape free of roughness
at the scale of the thermal energy~\cite{VaVe09}.
In any case, the equations defining a center-of-density
path are intrinsically more expensive computationally to solve
than those for the MFTP, because they require averaging
on quasi-committor isosurfaces $\hat{q}(\zeta) =$ constant
(in addition to conditional averages on collective variable isosurfaces
$\xi(x) = \zeta$ in full configuration space)
rather than merely determining a (local) maximum.


\section{A method}  \label{sec:meth}

As stated previously, computing $\hat{q}(\zeta)$ is not feasible.
Consequently, we derive a method,
which employs three uncontrolled approximations---a
controlled approximation being one that can be made arbitrarily
accurate with sufficient computational effort.
Subsection~\ref{sss:brownian}
approximates paths in collective variable space by those of Brownian dynamics;
Subsection~\ref{sss:mftp} assumes most paths lie in a tube where isocommittors are planar; and
Subsection~\ref{sss:mfep} assumes that on average the trajectories are parallel to the path.
The basic ingredients of much of this development are present
in the literature but scattered among several articles.
Here they are combined to produce equations from which we derive the MFTP.

\subsection{Brownian dynamics approximation of collective variable paths}\label{sss:brownian}

The probability density function (p.d.f.) for $\xi(x)$ is
\[ \rho_\xi(\zeta) = \langle\delta(\xi(x) - \zeta)\rangle
= \iint\delta(\xi(x) - \zeta)\rho(x, p)\rmd x\rmd p \]
where $\delta(\zeta) = \delta(\zeta_1)\delta(\zeta_2)\cdots\delta(\zeta_\nu)$.
Let $\langle\cdot\rangle_\zeta$
be the expectation for the conditional density $\rho(x, p|\xi(x)=\zeta)$:
$$\langle O(x)\rangle_\zeta
= \frac{\langle\delta(\xi(x)-\zeta)O(x)\rangle}%
{\langle\delta(\xi(x)-\zeta)\rangle}.$$

In Appendix~\ref{app:brown}
is an adaptation of an argument from Ref.~\cite{MFVC06}~(Sec.~III, A~and~B)
suggesting that as an approximation to $\hat{q}(\zeta)$,
we should seek a function $q(\zeta)$ that minimizes a certain
functional $I(q)$ that can be expressed in terms of collective
variables $\zeta$.
Define the free energy $F(\zeta)$ for coordinates $\zeta = \xi(x)$ by
\begin{equation}\label{L1}
 \mathit{const}_\xi\rme^{-\beta F(\zeta)} =
 \rho_\xi(\zeta) = \langle\delta(\xi(x) - \zeta)\rangle.
\end{equation}
Also define a proto-diffusion tensor $D$ by
\[ D(\zeta) = \frac{1}{2}\beta^{-1}\langle\xi_x(x)M^{-1}\xi_x(x)\T\rangle_\zeta. \]
(There is freedom in the scaling of $D$.
We use this freedom to make Eq.~(\ref{eq:brownian}) agree with an alternative
derivation of the Brownian dynamics, in which
one assumes instantaneous relaxation of the degrees of freedom
not represented by the collective variables.
The tensor $D(\zeta)$ fails to be a diffusion tensor because it is missing
a time scale factor.)
The functional is then
\begin{equation} \label{eq:varform}
I(q) = \mathit{const}_\xi
\int\rme^{-\beta F(\zeta)}\nabla q(\zeta)\T D(\zeta)\nabla q(\zeta)\rmd\zeta
\end{equation}
where the integral is over the transition region outside of $A_\xi$ and $B_\xi$
subject to $q(\zeta) = 0$ on the boundary of $A_\xi$ and
$q(\zeta) = 1$ on the boundary of $B_\xi$.

The corresponding Euler-Lagrange equation for $q(\zeta)$ is
the Smoluchowski (backward Kolmogorov) equation:
\begin{equation} \label{eq:smoluch}
  - \nabla\cdot \rme^{-\beta F(\zeta)}D(\zeta)\nabla q(\zeta) = 0,
\end{equation}
subject to $q(\zeta) = 0$ on the boundary of $A_\xi$ and
$q(\zeta) = 1$ on the boundary of $B_\xi$.

The function $q$ that satisfies the Smoluchowski equation
subject to the given boundary conditions can be shown
to be the exact committor function for paths $\zeta = \zeta(\tau)$
in collective variable space generated by the Brownian dynamics
\begin{equation}\label{eq:brownian}
\frac{\rmd}{\rmd\tau}\zeta =  - \beta D(\zeta)\nabla F(\zeta)
 + (\nabla\cdot D(\zeta))\T
 +\sqrt{2}D_{1/2}(\zeta)\eta(\tau)
\end{equation}
where $D_{1/2} D_{1/2}\T = D$
and $\eta(\tau)$ is a collection of standard white noise processes.
The fact that $\tau$ is an artificial time does not affect the committor.
In principle, the assumption $q(\zeta)\approx \hat{q}(\zeta)$
can be checked a posteriori
by comparing committor values of the Brownian dynamics
to the quasi-committor values of actual dynamics.

Reference~\cite{MFVC06}~(Sec.~III.C)
appears to suggest
that the Smoluchowski equation
uniquely specifies dynamics except for scaling of time:
If the Smoluchowski equation~(\ref{eq:smoluch})
is satisfied by committors $q(\zeta)$  for arbitrary
sets $A'_\xi$ and $B'_\xi$ in collective variable space,
then trajectories whose committor functions satisfy Eq.~(\ref{eq:smoluch})
must have paths that are those of the Brownian  dynamics.
Hence, paths in collective variable space
can be generated with the proper probabilities from the system of stochastic
differential equations.

\subsection{Last hitting-point distribution}\label{sss:lasthit}

Appendix~\ref{app:lasthit} considers the rate at which
reactive trajectories cross an arbitrary surface $\Sigma$
that separates collective variable space into two parts,
one containing $A_\xi$ and the other containing $B_\xi$.
The result given there is that the rate of the {\em last} crossing of
$\Sigma$ by reactive trajectories is given by the integral
\[ \int_\Sigma J(\zeta)\cdot \hat{n}(\zeta) dS_\zeta,\]
where $\hat{n}(\zeta)$ points to the side containing $B_\xi$, and
\[ J(\zeta) = \rho_\xi(\zeta)D(\zeta)\nabla q(\zeta) \]
is the last hitting-point flux.
The choice of the last hitting point to represent the point where a reactive
trajectory crosses an isocommittor is somewhat arbitrary.
Therefore, it is gratifying to know that the expression for $J(\zeta)$
also gives the net flux and the first hitting-point flux
of reactive trajectories.

The normal to an isocommittor is given by $\hat{n}(\zeta) = \nabla q(\zeta)/|\nabla q(\zeta)|$,
so the distribution of last hitting points on an isocommittor
is proportional to
\[ j(\zeta) = \rho_\xi(\zeta)
\nabla q(\zeta)\T D(\zeta)\nabla q(\zeta)/|\nabla q(\zeta)|.
\]

\subsection{Defining the path} \label{sss:defpat}

For computation it is convenient to label the isocommittors
with the path parameter $\arcl$.
In particular, denote by $\Sigma(\arcl)$, $0\leq\arcl\leq 1$,
the isocommittor passing through $\zeta = \pth(\arcl)$.
Write $\bar{q}(\arcl) = q(\pth(\arcl))$ and define $\sigma(\zeta)$ implicitly by
\begin{equation}
\label{eq:dcmp} q(\zeta) = \bar{q}(\sigma(\zeta)).
\end{equation}
In this way the committor $q(\zeta)$ is decomposed into
two independent parts:
one part $\sigma(\zeta)$ specifies the isocommittor label and the other part
$\bar{q}(\arcl)$ calibrates the isocommittors.
On an isocommittor $\Sigma(\arcl)$, the gradient
$\nabla q(\zeta) = \bar{q}_s(\sigma(\zeta)) \nabla \sigma(\zeta)$,
so the normal flux is
\begin{equation}\label{L2}
j(\zeta) =  \bar{q}_s(\sigma(\zeta)) \rho_\xi(\zeta)
\nabla \sigma(\zeta)\T D(\zeta)\nabla \sigma(\zeta)/|\nabla \sigma(\zeta)|
\end{equation}
(recalling that the subscript $s$ denotes differentiation $\mathrm{d}/\mathrm{d}s$).
Note that $\bar{q}_s$ contributes only a scale factor to $j(\zeta)$,
so the center of intensity on $\Sigma(\arcl)$ does not depend on $\bar{q}_s$.

Each point $\pth(\arcl)$ on the desired path maximizes
the last hitting-point flux $j(\zeta)$ on the isocommittor
$q(\zeta) = q(Z(s))$.
Hence, $\nabla j(Z(s))~\|~\nabla q(Z(s))$.
To keep the derivation independent of the calibration $\bar{q}(s)$,
introduce a vector $n(s)$, not necessarily normalized,
 such that $n(s)~\|~\nabla q(Z(s))$.
Hence,
\begin{equation}\label{eq:whatever}
 \nabla j(Z(s))~\|~n(s).
\end{equation}

\subsection{The localized tube assumption} \label{sss:locatube}

Assume there exists a tube connecting $A_\xi$ to $B_\xi$
such that
(i) on each isocommittor,
  regions of high $j(\zeta)$ are concentrated in the tube,
(ii) each isocommittor is nearly planar in the tube, and
(iii) $D(\zeta)$ is nearly constant on each isocommittor within the tube.
This scenario is illustrated in \ref{fig:comm} below.

\begin{figure}[htbp]
\begin{center}
\includegraphics[width=3.375in]{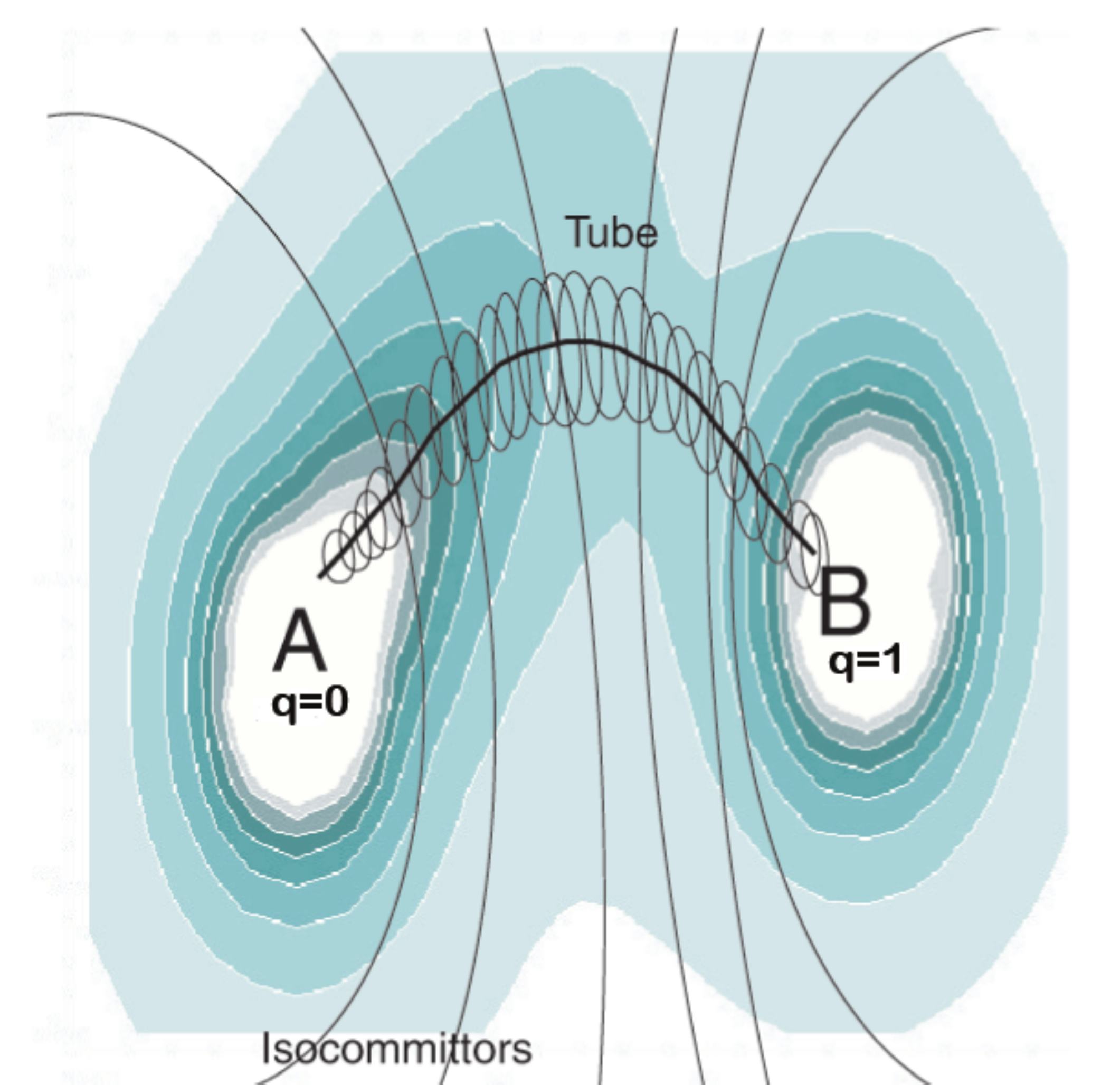}
\caption{ Shading indicates contours of free energy, thin curves denote
isocommittors, ellipses
enclose concentrations of crossing points from reactive trajectories, and
the thick curve is the center.}
\label{fig:comm}
\end{center}
\end{figure}

Exploit the localized tube assumption by
approximating the  isocommittor through $\pth(\arcl)$
as a plane $\Pi(\arcl)$ with normal $n(\arcl)$.
Hence, the isocommittor surface $\Sigma(\arcl):~\sigma(\zeta) = \arcl$
has the simple description of a hyperplane,
\begin{equation}\label{eq:planar}
\Pi(\arcl) :\quad n(\arcl)\cdot(\zeta - \pth(\arcl)) = 0.
\end{equation}
Also, approximate $D(\zeta)$ by
$\bar{D}(\sigma(\zeta))$
where $\bar{D}(s) \stackrel{\mathrm{def}}{=} D(Z(s))$.
These approximations (see Ref.~\cite{Vand06}~(Sec.~6.6.1))
are sufficient to define a practical method (see Ref.~\cite{EVa04}~(Sec.~12)).
The unknown direction vector $n(\arcl)$ is to be chosen to minimize
the integral $I(q)$ of Eq.~(\ref{eq:varform}) restricted to some tube.
For simplicity the boundary points $\pth(0)$ and  $\pth(1)$
can be moved to points in $A_\xi$ and $B_\xi$ that locally
minimize $F(\zeta)$.
In this way the problem of solving for a committor of many variables
is reduced to that of a one-dimensional calculation along the length of
the tube.

It remains to derive the condition that determines $Z(s)$.
This is done in Appendix~\ref{app:deriv},
where it is shown that the condition is
\[
 -\beta\nabla F(Z) + \frac{n_s}{n\T Z_s}~\|~n.
\]

\subsection{The maximum flux transition path}\label{sss:mftp}

Although the localized tube assumption is sufficient
for defining a practical method, the method would not be simple,
so we make an additional simplifying assumption:
Assume the flux $J(\zeta)$ points in the direction of the path so that
$ J(\pth(\arcl)) ~\|~ \pth_s(\arcl)$
or $ D(Z(s))\nabla q(Z(s)) ~\|~ Z_s(s)$, whence
\[ n(s)~\|~D(Z(s))^{-1}\pth_s(\arcl).\]
The result is a {\em maximum flux transition path}
\begin{equation}  \label{eq:mftp}
 - \beta\nabla F(Z)
 + \frac{(D(Z)^{-1}Z_s)_s}{Z_s\T D(Z)^{-1}Z_s} ~\|~D(Z)^{-1}Z_s.
\end{equation}
(The simplifying assumption is justified, for example,
if the probability is strongly peaked around the path,
resulting in most of the probability contained in a narrow tube
with a flux $J(\zeta)$ pointing in the direction of the tube and the path.)
This assumption is also made for the FTS method,
see Eq.~(14) of Ref.~\cite{RVME05} and Sec.~II.A. of Ref.~\cite{VaVe09}.
Geometrically, this condition means that
instead of having the free energy gradient vanish orthogonal to the path,
it is balanced by a ``centripetal'' force,
which reduces curvature and avoids cusps.

To express condition~(\ref{eq:mftp}) as an equation, write it as
$-\beta(Z_s\T D^{-1}Z_s)D\nabla F + D(D^{-1}Z_s)_s = \lambda Z_s$
where $\lambda$ is a scalar and premultiply by $Z_s\T$ to
obtain an expression for $\lambda$.
After eliminating $\lambda$, the equation becomes
\begin{equation}  \label{eq:one}
(I - \Pi)D(D^{-1}Z_s)_s = (I - \Pi)\beta(Z_s\T D^{-1}Z_s)D\nabla F
\end{equation}
where the projector
\[\Pi = Z_s Z_s\T/(Z_s\T Z_s).\]

Note that, if $D$ is constant, the limit $\beta\rightarrow 0$ for
Eq.~(\ref{eq:one}) gives a geodesic $Z_{\arcl\arcl} = 0$,
which is the desired result.

In the two-dimensional case with $D = I$, the Euclidean length of
$(I-\Pi)D(D^{-1}Z_s)_s/(Z_s\T D^{-1}Z_s)$ is exactly equal to the curvature,
which is defined to be the reciprocal of the radius of curvature.
To see this, note that this is true if we parameterize with (actual) arc length
and note also that the curvature term is independent of parameterization
(which can be checked analytically).

If we normalize the parameterization using $(Z_s\T Z_s)_s = 0$,
this implies $Z_s\T Z_{ss} = 0$ and
$\Pi Z_{ss} = 0$.
Adding this last equation to Eq.~(\ref{eq:one}) gives
a {\em nonsingular} second order ordinary differential equation for $Z(s)$:
\[Z_{ss}
 = (I - \Pi)\left(\beta(Z_s\T D^{-1}Z_s)D\nabla F + D_s D^{-1}Z_s\right).\]

Values obtained from constructing the path can be used
to calculate the free energy $F(Z(s))$ along the path,
\[F(Z(\arcl)) - F(Z(0)) = \int_0^\arcl
 \nabla F(Z(\arcl'))\T Z_s(\arcl')
 \rmd \arcl'.\]
However, $F(Z(s))$ is not a potential of mean force for the transition.

\subsection{The minimum free energy path}\label{sss:mfep}

The simplifying assumption of the preceding subsection, which is used to derive the MFTP,
is valid in the limit $\beta\rightarrow\infty$ in the Brownian dynamics approximation;
see Ref.~\cite{Vand06}~(Sec.~6.6) and Ref.~\cite{MFVC06}~(App.~A).
A more systematic derivation might therefore neglect the curvature term.
The result would be a {\em minimum free energy path}
\[Z_s ~\|~ - \beta D(Z)\nabla F(Z). \]
Each point $\zeta = Z(\arcl)$ on the MFEP
is a local minimum of $F(\zeta)$
in the hyper-plane orthogonal to $D(Z(s))^{-1}Z_s(\arcl)$.

One difference from an MFTP is that
an MFEP can have a cusp at an intermediate local minimum.
If the path passes sufficiently close to a local minimum $\zeta = \zeta_0$
of $F(\zeta)$, then for a short section of the path, $\zeta = Z(s)$, $a\leq s\leq b$,
a quadratic approximation to $F(\zeta)$ is accurate.
Assume $D =$ constant and
$F(\zeta) = \frac12(\zeta-\zeta_0)\T A(\zeta-\zeta_0) +$ constant,
where $A$ is symmetric positive definite.
The MFEP is then defined by $Z_s~\|~ -\beta DA(Z - \zeta_0)$.
Perform a change of variables, $Y = \beta^{-1/2}Q\T D_{1/2}^{-1}(Z - \zeta_0)$
where $Q\Lambda Q\T$ is a diagonalization of $D_{1/2}\T A D_{1/2}$.
The MFEP for $Y(s)$ is hence given by $Y_s~\|~ -\Lambda Y$.
For simplicity, suppose that $Y = [x, y]\T$, that $x(a) < 0 < x(b)$,
and that $\Lambda = \mathrm{diag}(\lambda, \mu)$ with $\lambda > \mu$.
The path is hence defined by $y_s/(\mu y) = x_s/(\lambda x)$,
which can be integrated to yield the path
\[ y = \left\{\begin{array}{l@{\quad}l}
(x/x(a))^{\mu/\lambda}y(a), & x(a)\leq x\leq 0, \\
(x/x(b))^{\mu/\lambda}y(b), & 0\leq x\leq x(b), \\
\end{array}\right.\]
which has a cusp at $x=0$.

The FTS method path is also likely to suffer from the presence of cusps,
because for a harmonic potential, the average position is the same
as the most probable position.

The presence of cusps undermines the localized tube assumption.
In particular,
the assumption of isocommittors being approximately planar breaks down
at a cusp.
This poses a difficulty when computing transition rates,
which depends on existence of isocommittors.
Additionally, cusps complicate the numerical approximation of paths.


\section{An algorithm}  \label{sec:alg}

An algorithm for calculating a transition path employs
a progression of four controlled approximations:
discretization of the path $\zeta = Z(\arcl)$
and the equations that define it;
a finite number of iterations for the solution of nonlinear discrete equations;
use of restraints for constrained sampling;
and finite sampling.

\subsection{Discretization}

The path $Z(\arcl)$, $0\leq \arcl\leq 1$, is approximated as a
piecewise polynomial with break points
$0 = \arcl_0 <\arcl_1 < \cdots <\arcl_J = 1$.
Here we choose a uniform mesh $\arcl = 0, \Delta\arcl, \ldots, 1$
and obtain the path by piecewise linear interpolation.
Thus the problem is reduced to determining {\em unknown} nodal values
$Z_j\approx Z(\arcl_j)$, $j = 0, 1, \ldots, J$,
each representing a replica of the system in a different configuration.

It is convenient for computation to use for the path parameter
$\arcl$ the arc length along the path divided by the total length of the path.
In such a case, $|\pth_s(\arcl)|$ is constant.
The arc length normalization becomes
\[
 |Z_{j+1} - Z_j| / \Delta\arcl
 = |Z_j - Z_{j-1}| / \Delta\arcl, \quad
 j = 1, 2, \ldots, J-1. \]

Condition~(\ref{eq:mftp}) is written as
\[
- \beta D\nabla F + \frac1{c}Z_{ss} - \frac{1}{c}D_s D^{-1}Z_s ~\|~ Z_s
\]
where $c = Z_s\T D^{-1}Z_s$.

This is discretized by the finite difference scheme
\[ (Z_s)_j~\|~g_j, \mbox{ where }
g_j \stackrel{\mathrm{def}}{=} - \beta D_j(\nabla F)_j + \frac{1}{c_j}\frac{Z_{j+1} - 2Z_j + Z_{j-1}}{\Delta s^2}
- \frac{1}{c_j}(D_s D^{-1}Z_s)_j
\]
and where
\begin{eqnarray}
c_j  & = & \frac12\Delta s^{-2}
 (\Delta_-Z_j\T D_j^{-1}\Delta_-Z_j + \Delta_+Z_j\T D_j^{-1}\Delta_+Z_j), \label{eq:disc1}
 \\
(D_s D^{-1}Z_s)_j & = & \frac12\Delta s^{-2}
 (\Delta_-D_j D_j^{-1}\Delta_-Z_j + \Delta_+D_j D_j^{-1}\Delta_+Z_j), \label{eq:disc2}
\end{eqnarray}
with
\[ \Delta_\pm D_j =\mp(D_j - D_{j\pm 1}), \mbox{ and }
\Delta_\pm Z_j =\mp(Z_j - Z_{j\pm 1}).
\]
We choose upwinded differencing for $(Z_s)_j$
based on the direction of the modified mean force $g_j$:
\begin{equation}{}\label{eq:upwind}
 (Z_s)_j = \left\{\begin{array}{l}
 (Z_j - Z_{j-1}) / \Delta s \quad\mbox{if } g_j\T(Z_j - Z_{j-1}) > 0,\\
 (Z_{j+1} - Z_j) / \Delta s \quad\mbox{if } g_j\T(Z_j - Z_{j+1}) > 0.
\end{array}\right.
\end{equation}
In the unlikely event that both conditions are satisfied, the choice
is dictated by the arc length normalization step
of the simplified string method to be discussed next.

For the MFEP, cusps can occur at some intermediate local minima,
requiring an adaptive mesh to resolve.

\subsection{Solution of nonlinear discrete equations}

A second component of the algorithm is an iterative method
for achieving rapid local convergence given a plausible initial guess.

Because of its simplicity and demonstrated effectiveness,
we adopt the semi-implicit simplified string method used in
Ref.~\cite{VaHe08}~(Eq.~(11)).
To determine a path, begin with an initial guess
and generate successive improvements
by alternating between moving the points of the curve $Z_j$
in the direction $g_j$ given by condition~(\ref{eq:mftp}) and reparameterizing.

The first step of each iteration is to solve the following equations for the $Z_j^\ast$:
\begin{eqnarray*}
\displaystyle\frac{Z_j^\ast-Z_j}{\tau^2} &=&
\displaystyle\frac{1}{c_j}\frac{Z_{j+1}^\ast - 2Z_j^\ast + Z_{j-1}^\ast}{\Delta s^2}
- \frac{1}{c_j}(D_s D^{-1}Z_s)_j - \beta D_j(\nabla F)_j, \ j =1, 2, \cdots, J-1, \\
\displaystyle\frac{Z_j^\ast-Z_j}{\tau^2} &=& - \beta D_j(\nabla F)_j, \  j =0, J,
\end{eqnarray*}
where $c_j$ and $(D_sD^{-1}Z_s)_j$ are given in Eqs.~(\ref{eq:disc1})~and~(\ref{eq:disc2}).
(The extra factor $\tau$ provides the time scale factor missing from $D$.)

Then the normalization adjustment is to
choose the $\{Z_j\}$ to be equidistant along the resulting curve:
\begin{eqnarray*}
 s_0^\ast &=& 0, s_j^\ast = s_{j-1}^\ast+|Z_j^\ast-Z_{j-1}^\ast|, \\
 Z^\ast(\arcl) &=& \mbox{piecewise linear interpolation of }
 \{(s_j^\ast/s_J^\ast, Z_j^\ast)\},\quad 0\leq \arcl\leq1, \\
 Z_j^\mathrm{new} &=& Z^\ast(j/J).
\end{eqnarray*}
It can be shown that if the semi-implicit simplified string method converges,
the resulting points $Z_j$ satisfy a nonstandard discretization of the
differential equation containing $\tau$ as a parameter.
In the limit $\tau\rightarrow 0$, the discretization becomes
upwinded differencing.

For large systems, targeted molecular dynamics~\cite{ScEK94} has been used
to get an initial path~\cite{GaYR09,HuOP09}.
Another potentially promising but quite different approach
is rigidity analysis~\cite{LZKT04}.

\subsection{Conditional averages}

Evaluation of $\nabla F$ and $D$ at break points
involves sampling on hyper-surfaces
$\{x : \xi(x) = Z_j\}$ of configuration space.

For calculating such conditional expectations,
the Dirac delta function $\delta(s)$ can be approximated by the p.d.f.\ of
a Gaussian
$\delta_\varepsilon(s)
=(2\pi\varepsilon^2)^{-1/2}\exp(-s^2/(2\varepsilon^2))$.
Note
$$\delta_\varepsilon(\xi(x)-\zeta)\rme^{-\beta U(x)}
=(2\pi\varepsilon^2)^{-\nu/2}\rme^{-\beta U(x;\zeta)}$$
where
\begin{equation}\label{eq:restrain}
U(x;\zeta)
=U(x)+\sum\limits_{i=1}^\nu u_i(x,\zeta_i), \quad\mbox{  and  }
u_i (x,\zeta_i) = \frac{1}{2\beta \varepsilon^2}(\xi_i(x)-\zeta_i)^2.
\end{equation}
Then,
$\langle O(x)\rangle_\zeta = \langle O(x)\delta_\varepsilon(\xi(x)-\zeta)\rangle/
\langle\delta_\varepsilon(\xi(x)-\zeta)\rangle$ is nothing but an average using $U(x; \zeta)$.
The effect is that of using restraining potentials instead of constraints.
These restraints should be as strong as possible
without restricting the step size used in the sampling.
From $\mathit{const}_\xi\exp(-\beta F(\zeta))=
\langle\delta_{\varepsilon}(\xi(x)-\zeta)\rangle$, we have
$$\nabla F(\zeta)
=-\frac{1}{\beta\varepsilon^2}\langle\xi(x)-\zeta\rangle_{\zeta}.$$

\subsection{Sampling}

We would like to estimate the statistical error of $Z_j^\ast$.
Ideally, we want the standard deviation of the estimate smaller than some given tolerance.
The major contribution to the sampling error of $Z_j^\ast$
comes from that of $(\nabla F)_j$, because of
the cancelation and subsequent multiplication by $\varepsilon^{-2}$. Thus,
we neglect the statistical error of $D_j$ in estimating the error of $g_j$.
So then, the statistical error of $Z_j^\ast$ comes from the sample average of
$\Delta_j = \beta D_j (\nabla F)_j^n,\ n=1,2,\ldots,N$, where
$N$ is the sample size.
The statistical error is defined by $(\max_{0\leq j\leq J} \mbox{error bar of $\Delta_j$})\tau^2$,
where an error bar is an estimate of 1 standard deviation.
Such an estimate can be obtained using block averaging as in
Ref.~\cite{FS02}~(Appendix~D.3).
In general, 32 blocks is a reasonable choice.

At each iteration, the configuration $x$ from the previous iteration
could be used to start the equilibration of the molecular dynamics.
Thus, it is necessary that values of $x$ be stored
such that $\xi(x) =  Z_j$, $j = 0, 1, \ldots, J$.
It is reasonable to expect less equilibration time is needed in later
iterations as the path converges.


\section{Numerical tests}  \label{sec:test}

\subsection{An artificial problem} \label{sec:arti}

As an example to illustrate our method, consider a problem finding the
MFTP and MFEP for the potential energy function
\begin{eqnarray*}
U (x, y) & = & -4 \exp(-4x^2 - (y - 2.75)^2 ) - 5 \exp(-(x - 1)^2 - (y-0.15)^2 ) \\
& & \mbox{} -5 \exp(-(x + 1)^2 - y^2 ) + 8 \exp(-x^2 - (y + 0.5)^2 ) + 0.001(x^4+y^4)
\end{eqnarray*}
where the energy unit is kcal/mol and the mass matrix $M$ has identical diagonal entries.
Unless specifically mentioned, the inverse temperature
$\beta^{-1} = 0.59595$\, kcal/mol, corresponding to 300\,K.
In particularly, we take collective variables $\zeta = \xi(x,y) = (x,y)$.
In this case, the MFEP becomes a minimum energy path (MEP).
Alternatively, an MEP can be considered as an MFEP
for which we have an accurate estimate of $F(\zeta)$.

\begin{figure}[htbp]
\begin{center}
\includegraphics[width=4.0in]{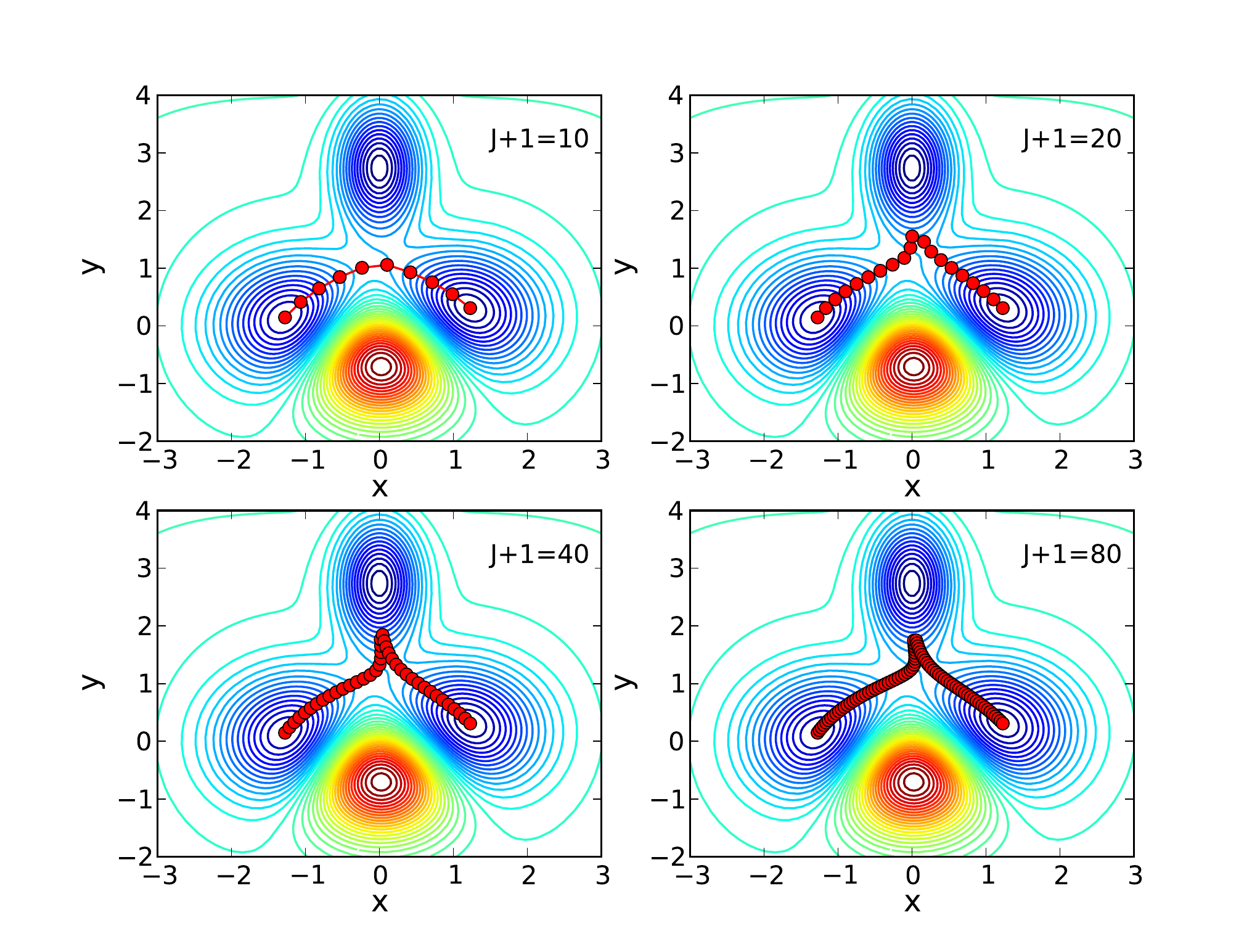}
\caption{Minimum energy path obtained using the simplified string method.
The initial path is the straight line between $(-1,0)$ and $(1,0)$.
The path is discretized into $J+1$ images.
Four figures are generated using $J+1=10$, $20$, $40$, $80$ images,
respectively.}
\label{fig:mep}
\end{center}
\end{figure}

In \ref{fig:mep}, we show an MEP connecting two local minima through
the third local minimum.
The MEP has a cusp at the intermediate minimum.
The MEPs are computed using the simplified string method
with piecewise linear interpolation and equal arc length normalization.
The time step $\tau^2 = 0.01$.
The iteration is stopped if $d<0.005$,
where $d= \max_{0\leq j\leq J}\tau^{-2}|Z_j^{\mathrm{new}}-Z_j|$.
From the figure, we can see that the cusp is missing if the
number of images ($J +1 = 10$) is
too small.
Also, the MEP does not go through the intermediate local minimum
as it should, even with many images ($J+1=80$).

A calculation (not shown here) similar to that for \ref{fig:mep} was done for the MFTP.
The MFTP is calculated using the semi-implicit simplified string method described in Sec.~\ref{sec:alg}.
The MFTP can be resolved using a relatively small set of images,
for example, the MFTP calculated by only 10 images ($J=9$)
is almost indistinguishable from the one calculated using 80 images ($J=79$).
The MFTP avoids the cusp problem.

\begin{figure}[htbp]
\begin{center}
\includegraphics[width=3.375in]{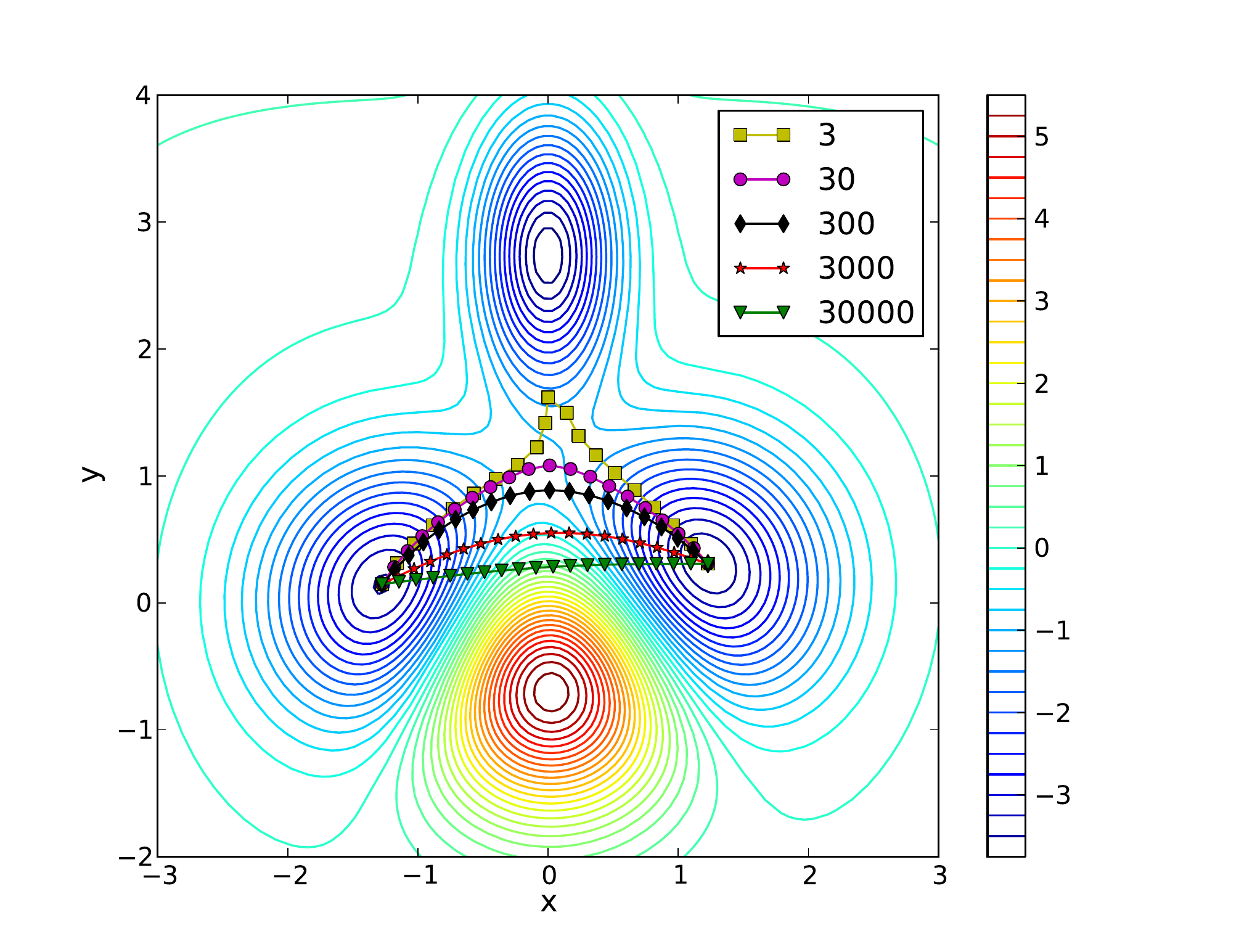}
\caption{Maximum flux transition path obtained using the semi-implicit
simplified string method.
Here we used the same initial path and the same stopping criterion
for convergence as for \ref{fig:mep}.
The MFTPs are generated using 20 images at 3\,K, 30\,K, 300\,K, 3000\,K,
30000\,K, respectively (which roughly correspond to $\beta^{-1} = 0.006, 0.06, 0.6, 6, 60$ kcal/mol.)
The contour lines are separated by 0.25\,kcal/mol.}
\label{fig:mftp}
\end{center}
\end{figure}

The MFTP generates different paths at different temperature.
\ref{fig:mftp}  shows MFTPs at 3\,K, 30\,K, 300\,K, 3000\,K, 30000\,K, respectively.
It is clear that the MFTP is close to the MEP at low temperature
(3\,K) and is close to a straight line at high temperature (30000\,K),
which is what we expect.

An FTS method path is expected to be similar to an MFEP for this example.

\subsection{Phi, psi for alanine dipeptide in vacuum}

For comparison with the MFEP, we study alanine dipeptide
at 300\,K in vacuum~\cite{MFVC06}.
We compare the MFEP and
the MFTP with two dihedral angles $\phi$ and $\psi$ as collective variables.
All simulations were performed using the CHARMM simulation
program~\cite{BBMN09,BBOSSK83}
and the full-atom representation of the molecule in the
CHARMM force field~\cite{MBBD98,MFB04}.
Langevin dynamics with friction coefficient $10.0 \mbox{ ps}^{-1}$
and time step 1.0\,fs was used.
For the calculation of $\nabla F$ and $D$,
harmonic potentials as in Eq.~(\ref{eq:restrain}) were added involving the dihedral
angles $\phi$ and $\psi$ with force constant 1000 kcal/(mol rad$^2$) (corresponding
to $\varepsilon = 1^\circ$).

The initial path in collective variable space is a straight line between
two points in $\phi-\psi$ space.
The path is discretized into $J+1$ images.
The configuration of alanine dipeptide at each image along the initial
path is built using the IC module in CHARMM with dihedral angles fixed at the
interpolated values.
Then follow 1000 steps of minimization and 50,000 steps of heating before the iteration starts.
Each iteration of the path involves 50,000 steps of equilibration and 500,000 steps of sampling.
The configuration at the final step of sampling in the previous iteration
is used as the initial configuration for the equilibration in the next iteration.

\begin{figure}[htbp]
\begin{center}
\includegraphics[width=3.375in]{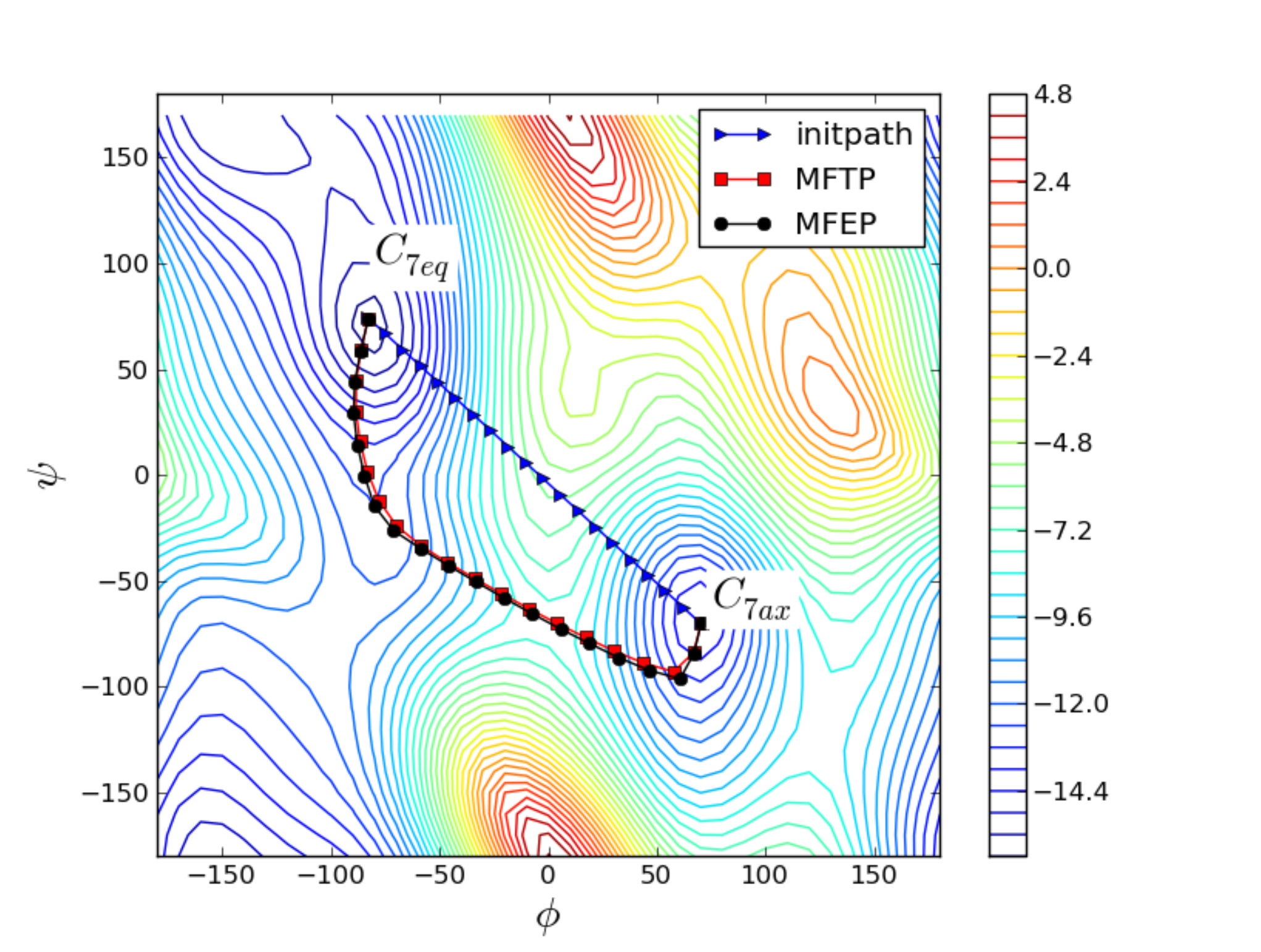}
\caption{ Maximum flux transition path and minimum free energy
path from $C_{7\mathrm{eq}}$ to $C_{7\mathrm{ax}}$ for alanine dipeptide
in vacuum at 300\,K.
Triangles are images of the initial path;
rectangles are the images of the maximum flux transition path;
and circles are the images of the minimum free energy path.
The contours are those for the zero-temperature free energy (adiabatic energy).
The contour lines are separated by 0.6\,kcal/mol.}
\label{fig:ad_nocusp}
\end{center}
\end{figure}

We begin by comparing the MFTP and MFEP from $C_{\mathrm{7eq}}$ to $C_{\mathrm{7ax}}$.
The MFEP is calculated using the simplified string method
with linear interpolation between images and equal arc length normalization.
The MFTP is calculated using the semi-implicit simplified string method.
In \ref{fig:ad_nocusp}, the initial path is the straight line between $(-83.2^\circ, 74.5^\circ)$
and $(70^\circ, -70^\circ)$, which were determined as $C_{\mathrm{7eq}}$ and $C_{\mathrm{7ax}}$
in Ref.~\cite{MFVC06}.
The path is discretized into 20 images.
The time step $\tau^2 = 0.16$ in CHARMM time units squared, or $\tau^2 = (19.56\, \mbox{fs})^2$.
The statistical error estimated by block averaging using 32 blocks is $\pm0.00577^\circ$.
The iteration is stopped if $d<0.02$.
(The tolerance value should be chosen properly since the statistical error will eventually dominate
the other errors so that $d$ fluctuates about a positive number.)
It takes 34 and 31 iterations to converge for the MFTP and MFEP, respectively.
The computational cost for two methods is comparable.
The path calculated for this problem by the FTS method using the
CHARMM force field is given in Figure~5 of Ref.~\cite{RVME05}.

\begin{figure}[htbp]
\begin{center}
\includegraphics[width=3.375in]{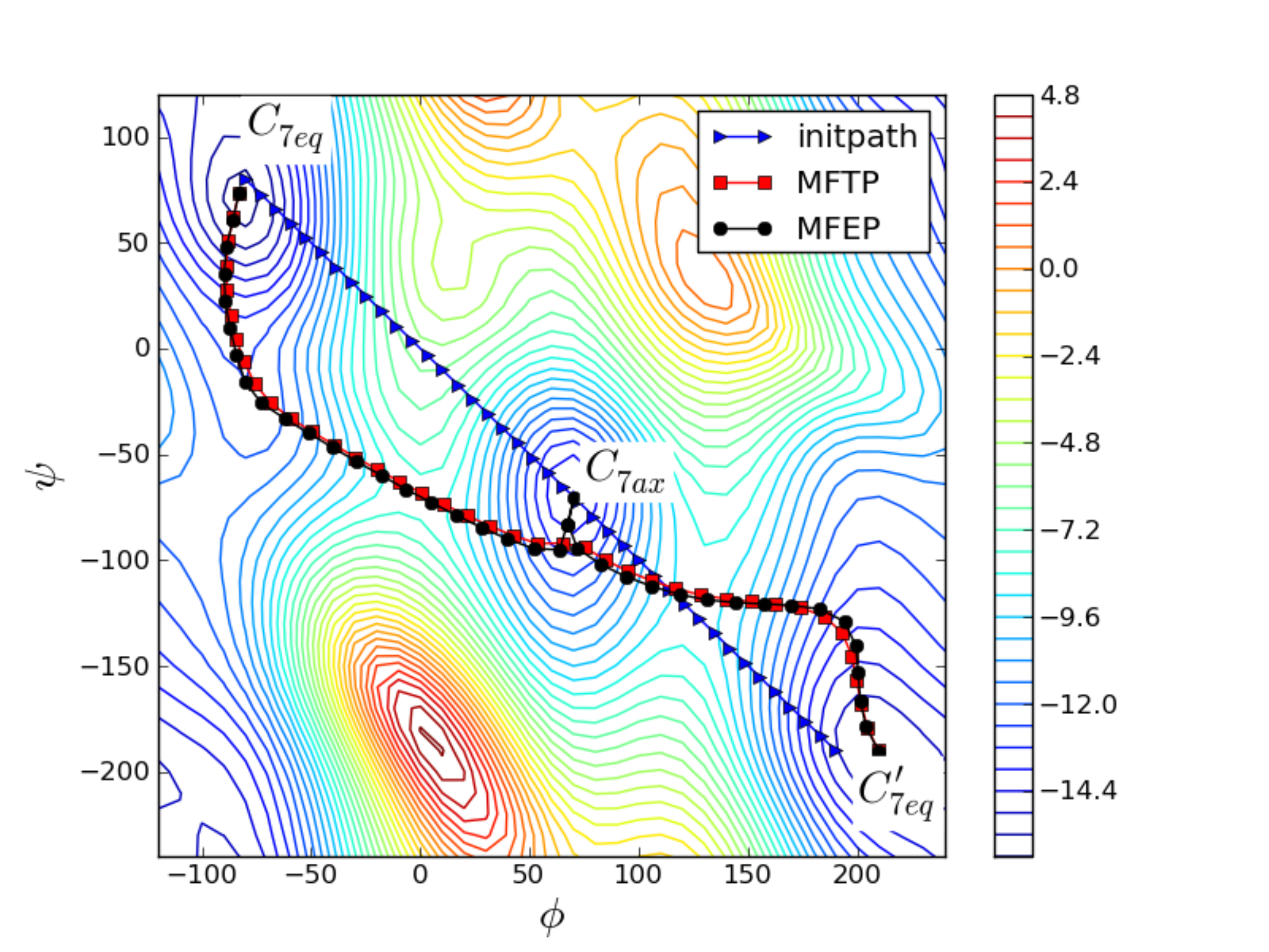}
\caption{Maximum flux transition path and minimum free energy
path for alanine dipeptide from $C_{\mathrm{7eq}}$ to $C'_{\mathrm{7eq}}$
passing by $C_{\mathrm{7ax}}$ in vacuum at 300\,K.
The figure is generated using 40 images.
Triangles are the images for the initial path;
rectangles are the images of the maximum flux transition path;
and circles are the images of the minimum free energy path.
The contours are those for the zero-temperature free energy.
The contour lines are separated by 0.6\,kcal/mol.}
\label{fig:ad_cusp}
\end{center}
\end{figure}

Next we compare the MFTP and MFEP from $C_{\mathrm{7eq}}$ to $C'_{\mathrm{7eq}}$.
In particular, we calculate the transition path $C_{\mathrm{7eq}}$--$C_{\mathrm{7ax}}$--$C'_{\mathrm{7eq}}$,
in which $C_{\mathrm{7ax}}$ serves as an intermediate metastable state.
The initial path is taken to be the straight line between ($-80^\circ$, $80^\circ$)
and ($190^\circ$, $-190^\circ$).
\ref{fig:ad_cusp} shows the MFTP and MFEP generated using 40 images.
The time step $\tau^2 = (19.56\,\mbox{fs})^2$.
The iteration is stopped if $d<0.02$.
It takes 35 and 44 iterations for the MFTP and MFEP to converge, respectively.
It is evident that the MFTP is more efficient than the MFEP in this case.

\subsection{Phi, psi for alanine dipeptide in solution}

We also test our method for alanine dipeptide solvated
in explicit water.
Again, the backbone dihedrals $\phi$ and $\psi$  are used as
collective variables to describe the transition.
The initial paths are straight lines connecting two points among
$(-77^\circ, 138^\circ)$, $(55^\circ, 48^\circ)$, $(60^\circ, -72^\circ)$,
and $(-77^\circ, -39^\circ)$ in $(\phi, \psi)$-space.

For preparing the simulation, each starting structure for alanine
dipeptide with constrained $\phi$ and $\psi$ angels is solvated
in a  $(20\times18\times15)$\,\AA$^3$ box with 191 TIP3~\cite{JCMIK83}
water molecules and equilibrated for 50,000\,ps.
The molecular dynamics are carried out with the CHARMM program
under the CHARMM22 force field.
Periodic boundary conditions are used and
the electrostatic interactions are treated with the particle mesh Ewald
method~\cite{EPBD95}. 
The system is simulated at a constant pressure of 1.0\,atm
and a constant temperature 300\,K with the algorithm based on
Hoover's methods.
We use a 1-fs timestep with the SHAKE~\cite{RCB77} algorithm
to keep all bonds involving hydrogen atoms at fixed lengths.

\begin{figure}[htbp]
\begin{center}
\includegraphics[width=3.375in]{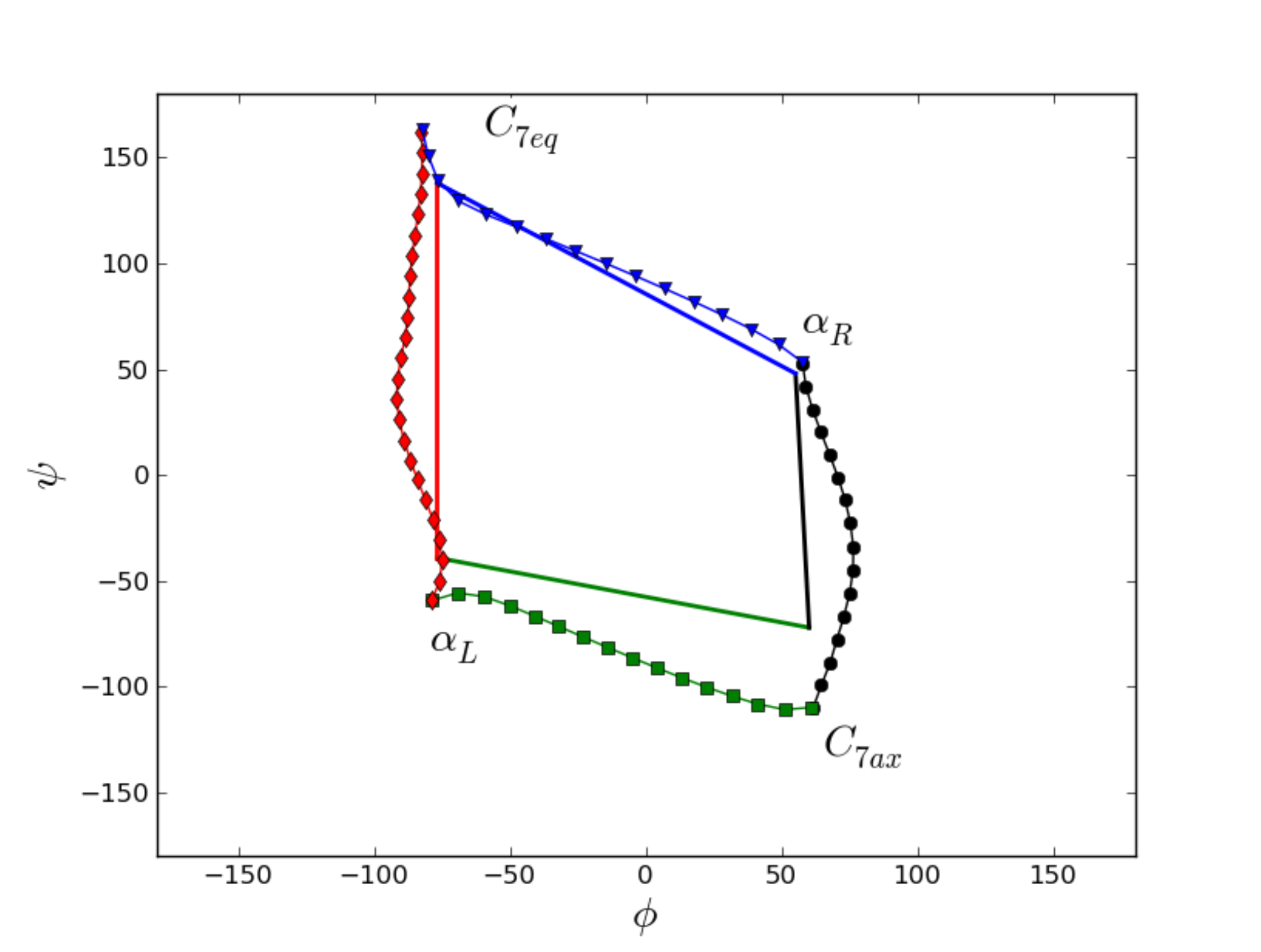}
\caption{Maximum flux transition paths for alanine dipeptide in solution.
The transition paths are calculated by the semi-implicit simplified string method
with the nearby straight lines as initial paths. }
\label{fig:solvad}
\end{center}
\end{figure}

In \ref{fig:solvad}, four MFTPs are calculated using the semi-implicit simplified
string method.
Each iteration involves 50,000 steps of equilibration and 500,000 steps
of sampling.
The transition paths are the result of 50 cycles of iteration.
The path $C_{\mathrm{7eq}}$--$\alpha_L$ calculated for this problem
by the FTP method using the CHARMM force field is given
in Figure~12 of Ref.~\cite{RVME05}.
The MFTP is similar to the FTS method path.

\section*{Acknowledgement}
This material is based upon work supported by grant R01GM083605
from the National Institute of General Medical Sciences, award A5286056128
from the University of Minnesota,
and by a 2007 Purdue Research Foundation Special Incentive Research Grant.
We would like to thank Carol Post for the collaboration that nurtured
this work.
Also, thanks to He Huang for
an initial implementation of the string method and an
early demonstration of cusps for alanine dipeptide,
and to Voichita Dadarlat for \ref{fig:comm}.
Additionally, thanks to Eric Vanden-Eijnden for helpful information about
transition path methods and theory,
and for suggestions that improved the original manuscript.
Finally, thanks to the Center for Biological Physics
at Arizona State University
and the Institute for Mathematics and Its Applications at the University
of Minnesota for providing environments that facilitated this work.


\appendix

\section{Derivation of Brownian dynamics approximation} \label{app:brown}

The quasi-committor is related to
a full phase-space committor in Ref.~\cite{Vand06}~(Sec.~6.~2)
$q^\ast$
defined as follows:
$$q^\ast(x, p)
= \Pr(X(t) \mbox{ reaches } B \mbox{ before } A
~|~X(0) = x, P(0)=p).$$
Note that $q^\ast(x, p) = 0$ or $1$,
because the dynamical equation is deterministic.
By definition, the quasi-committor
$\hat{q}(\zeta) = \langle q^\ast(x, p)\rangle_\zeta$.

It is not difficult to show that $\hat{q}(\xi(x))$ approximates $q^\ast(x, p)$
in the sense that it minimizes
$\langle|q(\xi(x)) - q^\ast(x, p)|^2\rangle$ over all $q(\zeta)$.
However, this is not useful for determining $\hat{q}(\zeta)$
because $q^\ast(x,p)$ is too costly to compute.
On the other hand, it is possible to find a best approximation to
$q^\ast(x,p)$ in another sense.
Because $q^\ast$ is constant on a trajectory, we have
\[0 = \frac{\rmd}{\rmd t}q^\ast(X(t), P(t)) = (Lq^\ast)(X(t), P(t))
\quad\mbox{where }L = (M^{-1}p)\cdot\nabla_x - U_x\cdot\nabla_p.\]
Consequently, $q^\ast$ satisfies the stationary Liouville equation
$$Lq^\ast=0, \qquad
q^\ast=0 \mbox{ on } A, \quad q^\ast=1 \mbox{ on } B.$$
Since we do know $L q^\ast = 0$, we seek instead an approximation
$q$ that minimizes $I(q) = \langle |L(q(\xi(x))-q^\ast(x,p))|^2\rangle$,
a standard tactic in numerical analysis.
As shown in Sec.~III.B of Ref.~\cite{MFVC06}, this simplifies to
\[ I(q) = \frac1{\beta}\langle|M^{-1/2}\nabla_x q(\xi(x))|^2\rangle,\]
which is to be as small as possible.
A low value for $I(q)$ is attained by having $q(\zeta)$ increase
monotonically from the value 0 on $A_\xi$ to the value 1 on $B_\xi$,
which is consistent with the prescription given earlier that $\xi(x)$ be chosen
so that the quasi-committor has no local minima or maxima outside of
$A_\xi$ and $B_\xi$.

The functional $I(q)$ can be expressed in terms of collective variables $\zeta$
as given by Eq.~(\ref{eq:varform}) and shown in in Eq.~(15) of Ref.~\cite{MFVC06}.

\section{Derivation of lasting hitting-point distribution}\label{app:lasthit}

The proof of Proposition~5 in Ref.~\cite{EVa06}~(p.~158)
analyzes the flux of reactive trajectories.
The flux $J(\zeta)$ gives the rate at which such trajectories
cross an arbitrary surface $\Sigma$
that separates collective variable space into two parts,
one containing $A_\xi$ and the other containing $B_\xi$
via the integral $\int_\Sigma J(\zeta)\cdot \hat{n}(\zeta) dS_\zeta$
where $\hat{n}(\zeta)$ points to the side containing $B_\xi$.
The proof actually looks not at all crossings but
only those occurring
within a vanishingly small time interval before the last
crossing---see Eq.~(50) of Ref.~\cite{EVa06}.
Therefore, it considers the net flux only in this limiting sense.
As the length of the time interval $\tau\rightarrow 0$, the positions
of these crossings all converge to the position of the last crossing.
So, indeed, one gets the flux of the last hitting point
from Proposition~5 of Ref.~\cite{EVa06}.
The result given in Ref.~\cite{EVa06}~(Eq.~(39)),
as well as in Ref.~\cite{MeSV06}~(Eq.~(6), Eq.~(A12)),
and Ref.~\cite{Vand06}~(Eq.~(62)),
is that the last hitting-point flux for reactive trajectories is
$ J(\zeta) = \rho_\xi(\zeta)D(\zeta)\nabla q(\zeta)$.
(Proposition~4 of Ref.~\cite{EVa06}
does not apply to the infinitely damped case of Langevin dynamics.)
That the expression for $J(\zeta)$
also gives the net flux of reactive trajectories is Eq.~(32) of Ref.~\cite{Vand06}.
Also, the formula for $j(\zeta)$ in Sec.~\ref{sss:lasthit} agrees in the special case $D = I$ with that for the {\em first}
hitting point distribution given
in Ref.~\cite{VVCE08}~(Appendix~B).
Last and first are the same for reversible dynamics like Brownian dynamics.
Finally, there is an example in Metzner, Sch\"utte, and Vanden-Eijnden (2006)
section III.C, where it is suggested to use $\hat{n}\cdot J$.

\section{Derivation of the maximum flux condition}  \label{app:deriv}

We have from (\ref{L1}) and (\ref{L2})
that the normal flux is
\begin{equation}\label{eq:deriv1}
j(\zeta)
= \mathit{const}_\xi\rme^{-\beta F(\zeta)}\bar{q}_s(\sigma(\zeta))
\nabla \sigma(\zeta)\T D(\zeta)\nabla \sigma(\zeta)/|\nabla \sigma(\zeta)|
\end{equation}
where $\sigma(\zeta)$ is defined implicitly by $q(\zeta) = \bar{q}(\sigma(\zeta))$.
And for each point $Z(s)$ on the desired path,
the condition to be satisfied (\ref{eq:whatever}) is $\nabla j(Z(s))~\|~n(s)$.
We also approximate $D(\zeta)$ by $\bar{D}(\sigma(\zeta))$
where $\bar{D}(s) \stackrel{\mathrm{def}}{=} D(Z(s))$.
Furthermore, the assumption~(\ref{eq:planar})
that isocommittors are planar
implies
\begin{equation}\label{eq:deriv2}
 n(\sigma(\zeta))\cdot(\zeta - Z(\sigma(\zeta))) = 0.
\end{equation}

Differentiating Eq.~(\ref{eq:deriv2}) w.r.t.\ $\zeta$, we get
\[(n_s(\sigma)\cdot(\zeta - Z(\sigma)) - n(\sigma)\cdot Z_s(\sigma))
\nabla\sigma + n(\sigma) = 0,\]
where the argument $\zeta$ of $\sigma$ has been omitted, whence
\begin{equation}\label{eq:deriv3}
\nabla\sigma = (n(\sigma)\cdot Z_s(\sigma) - n_s(\sigma)\cdot
(\zeta - Z(\sigma)))^{-1} n(\sigma).
\end{equation}
Substituting Eq.~(\ref{eq:deriv3}) into Eq.~(\ref{eq:deriv1})
and replacing $D(\zeta)$ by $\bar{D}(\sigma(\zeta))$,
the normal flux becomes
\[
j(\zeta) = \varphi(\sigma(\zeta), \zeta),
\]
where
 \[\varphi(s, \zeta) = \mathit{const}_\xi\rme^{-\beta F(\zeta)}
\bar{q}_s(s) (n(s)\cdot Z_s(s) - n_s(s)\cdot (\zeta - Z(s))^{-1} n(s)\T \bar{D}(s) n(s)/|n(s)|.\]
Note that
 \[\frac{\nabla_\zeta\varphi}{\varphi}
 = -\beta\nabla F + \frac{n_s}{n\cdot Z_s - n_s\cdot(\zeta - Z)},\]
and
 \[\left.\frac{\nabla_\zeta\varphi}{\varphi}\right|_{\zeta=Z}
 = -\beta\nabla F(Z) + \frac{n_s}{n\T Z_s}.\]
Thus, we have
\[
 \frac{\nabla j}{j} = \frac{(\nabla_\zeta\varphi)(\sigma(\zeta), \zeta)}{\varphi(\sigma(\zeta), \zeta)}
    + \frac{\varphi_s(\sigma(\zeta), \zeta)}{\varphi(\sigma(\zeta), \zeta)}\nabla\sigma(\zeta),
\]
and
\[\left.\frac{\nabla j}{j} \right|_{\zeta=Z}
 = -\beta\nabla F(Z) + \frac{n_s}{n\T Z_s}
 + \frac{\varphi_s(s, Z)n}{\varphi(s, Z)n(s)\T Z_s(s)}.\]
Hence, the condition is that
\[
 -\beta\nabla F(Z) + \frac{n_s}{n\T Z_s}~\|~n.
\]

\bibliography{paper}

\begin{thebibliography}{10}

\bibitem{BMMN83}
M.~Berkowitz, J.~D. Morgan, J.~A. Mc{C}ammon, and S.~H. Northrup.
\newblock Diffusion-controlled reactions: A variational formula for the optimum
  reaction coordinate.
\newblock {\em J. Chem.\ Phys.}, 79:5563--5565, 1983.

\bibitem{BCDG02}
P.~G. Bolhuis, D.~Chandler, C.~Dellago, and P.~L. Geissler.
\newblock Transition path sampling: Throwing ropes over rough mountain passes,
  in the dark.
\newblock {\em Annu.\ Rev.\ Phys.\ Chem.}, 53:291--318, 2002.

\bibitem{BoDC00}
P.~G. Bolhuis, C.~Dellago, and D.~Chandler.
\newblock Reaction coordinates of biomolecular isomerization.
\newblock {\em PNAS}, 97:5877--5882, 2000.

\bibitem{BBMN09}
B.~R. Brooks, C.~L. {Brooks III}, A.~D. {MacKerell Jr.}, L.~Nilsson, R.~J.
  Petrella, B.~Roux, Y.~Won, G.~Archontis, C.~Bartels, S.~Boresch, A.~Caflisch,
  L.~Caves, Q.~Cui, A.~R. Dinner, M.~Feig, S.~Fischer, J.~Gao, M.~Hodoscek,
  W.~Im, K.~Kuczera, T.~Lazaridis, J.~Ma, V.~Ovchinnikov, E.~Paci, R.~W.
  Pastor, C.~B. Post, J.~Z. Pu, M.~Schaefer, B.~Tidor, R.~M. Venable, H.~L.
  Woodcock, X.~Wu, W.~Yang, D.~M. York, and M.~Karplus.
\newblock {CHARMM}: The biomolecular simulation program.
\newblock {\em J. Comput.\ Chem.}, 30:1545--1614, 2009.

\bibitem{BBOSSK83}
B.~R. Brooks, R.~E. Bruccoleri, B.~D. Olafson, D.~J. States, S.~Swaminathan,
  and M.~Karplus.
\newblock {CHARMM}: A program for macromolecular energy, minimization, and
  dynamics calculations.
\newblock {\em J. Comput.\ Chem.}, 4:187--217, 1983.

\bibitem{ERV07}
W.~E, W.~Ren, and E.~Vanden-Eijnden.
\newblock Simplified and improved string method for computing the minimum
  energy paths in barrier-crossing events.
\newblock {\em J. Chem.\ Phys.}, 126:164103, 2007.

\bibitem{EVa04}
W.~E and E.~Vanden-Eijnden.
\newblock Metastability, conformation dynamics, and transition pathways in
  complex systems.
\newblock In S.~Attinger, editor, {\em Multiscale Modelling And Simulation},
  pages 35--68. Springer Verlag, 2004.

\bibitem{EVa06}
W.~E and E.~Vanden-Eijnden.
\newblock Towards a theory of transition paths.
\newblock {\em J. Stat.\ Phys.}, 123:503--523, 2006.

\bibitem{EPBD95}
U.~Essmann, L.~Perera, M.~L. Berkowitz, T.~Darden, H.~Lee, and L.~Pederson.
\newblock A smooth particle mesh {E}wald method.
\newblock {\em J. Chem.\ Phys.}, 103:8577--8593, 1995.

\bibitem{FS02}
D.~Frenkel and B.~Smit.
\newblock {\em Understanding Molecular Simulation: From Algorithms to
  Applications}.
\newblock Academic Press, 2002.

\bibitem{GaYR09}
W.~Gan, S.~Yang, and B.~Roux.
\newblock Atomistic view of the conformational activation of {S}rc kinase using
  the string method with swarms-of-trajectories.
\newblock {\em Biophys.\ J.}, 97:L8--L10, 2009.

\bibitem{HuOP09}
H.~Huang, E.~Ozkirimli, and C.~B. Post.
\newblock Comparison of three perturbation molecular dynamics methods for
  modeling conformational transitions.
\newblock {\em J. Chem.\ Theory Comput.}, 5:1304--1314, 2009.

\bibitem{HuSt97}
S.~Huo and J.~E. Straub.
\newblock The {M}ax{F}lux algorithm for calculating variationally optimized
  reaction paths for conformational transitions in many body systems at finite
  temperature.
\newblock {\em J. Chem.\ Phys.}, 107:5000--5006, 1997.

\bibitem{JoMJ98}
H.~J\'onsson, G.~Mills, and K.~W. Jacobsen.
\newblock Nudged elastic band method for finding minimum energy paths of
  transitions.
\newblock In B.~J. Berne, G.~Ciccotti, and D.~F. Coker, editors, {\em Classical
  and Quantum Dynamics in Condensed Phase Simulations}, page 385. World
  Scientific, Singapore, 1998.

\bibitem{JCMIK83}
W.~L. Jorgensen, J.~Chandrasekhar, J.~D. Madura, R.~W. Impey, and M.~L. Klein.
\newblock Comparison of simple potential functions for simulating liquid water.
\newblock {\em J. Chem.\ Phys.}, 79:926--935, 1983.

\bibitem{LZKT04}
M.~Lei, M.~I. Zavodszky, L.~A. Kuhn, and M.~F. Thorpe.
\newblock Sampling protein conformations and pathways.
\newblock {\em J. Comput.\ Chem.}, 25:1133--1148, 2004.

\bibitem{MaDi05}
A.~Ma and A.~R. Dinner.
\newblock Automatic method for identifying reaction coordinates in complex
  systems.
\newblock {\em J. Phys.\ Chem.\ B}, 109:6769--6779, 2005.

\bibitem{MBBD98}
A.~D. {MacKerell Jr.}, D.~Bashford, M.~Bellott, R.~L. {Dunbrack Jr.}, J.~D.
  Evanseck, M.~J. Field, S.~Fischer, J.~Gao, H.~Guo, S.~Ha, D.~Joseph-McCarthy,
  L.~Kuchnir, K.~Kuczera, F.~T.~K. Lau, C.~Mattos, S.~Michnick, T.~Ngo, D.~T.
  Nguyen, B.~Prodhom, W.~E. {Reiher, III}, B.~Roux, M.~Schlenkrich, J.~C.
  Smith, R.~Stote, J.~Straub, M.~Watanabe, J.~Wi\'{o}rkiewicz-Kuczera, D.~Yin,
  and M.~Karplus.
\newblock All-atom empirical potential for molecular modeling and dynamics
  studies of proteins.
\newblock {\em J. Phys.\ Chem.\ B}, 102:3586--3616, 1998.

\bibitem{MFB04}
A.~D. {Mackerell Jr.}, M.~Feig, and C.~L. {Brooks III}.
\newblock Extending the treatment of backbone energetics in protein force
  fields: Limitations of gas-phase quantum mechanics in reproducing protein
  conformational distributions in molecular dynamics simulations.
\newblock {\em J. Comput.\ Chem.}, 25:1400--1415, 2004.

\bibitem{MFVC06}
L.~Maragliano, A.~Fischer, E.~Vanden-Eijnden, and G.~Ciccotti.
\newblock String method in collective variables: Minimum free energy paths and
  isocommittor surfaces.
\newblock {\em J. Chem.\ Phys.}, 125:024106, 2006.

\bibitem{MeSV06}
P.~Metzner, C.~Sch\"utte, and E.~Vanden-Eijnden.
\newblock Illustration of transition path theory on a collection of simple
  examples.
\newblock {\em J. Chem.\ Phys.}, 125:084110, 2006.

\bibitem{OlEl96}
R.~Olender and R.~Elber.
\newblock Calculation of classical trajectories with a very large time step:
  Formalism and numerical examples.
\newblock {\em J. Chem.\ Phys.}, 105:9299--9315, 1996.

\bibitem{Onsa38}
L.~Onsager.
\newblock Initial recombination of ions.
\newblock {\em Phys.\ Rev.}, 54:554--557, 1938.

\bibitem{OzPo06}
E.~Ozkirimli and C.~B. Post.
\newblock {S}rc kinase activation: A switched electrostatic network.
\newblock {\em Protein Sci.}, 15:1051--1062, 2006.

\bibitem{OYMP08}
E.~Ozkirimli, S.~S. Yadav, T.~W. Miller, and C.~B. Post.
\newblock An electrostatic network and long-range regulation of {S}rc kinases.
\newblock {\em Protein Sci.}, 17:1871--1880, 2008.

\bibitem{PaSR08}
A.~C. Pan, D.~Sezer, and B.~Roux.
\newblock Finding transition pathways using the string method with swarms of
  trajectories.
\newblock {\em J. Phys.\ Chem.\ B}, 112:3432--3440, 2008.

\bibitem{PSLS03}
S.~Park, M.~K. Sener, D.~Lu, and K.~Schulten.
\newblock Reaction paths based on mean first-passage times.
\newblock {\em J. Chem.\ Phys.}, 119:1313--1319, 2003.

\bibitem{NAMD}
J.~C. Phillips, R.~Braun, W.~Wang, J.~Gumbart, E.~Tajkhorshid, E.~Villa,
  C.~Chipot, R.~D. Skeel, L.~Kal\'{e}, and K.~Schulten.
\newblock Scalable molecular dynamics with {NAMD}.
\newblock {\em J. Comput.\ Chem.}, 26:1781--1802, 2005.

\bibitem{RVME05}
W.~Ren, E.~Vanden-Eijnden, P.~Maragakis, and W.~E.
\newblock Transition pathways in complex systems: Application of the
  finite-temperature string method to the alanine dipeptide.
\newblock {\em J. Chem.\ Phys.}, 123:134109, 2005.

\bibitem{RCB77}
J.-P. Ryckaert, G.~Ciccotti, and H.~J.~C. Berendsen.
\newblock Numerical integration of the cartesian equations of motion of a
  system with constraints: molecular dynamics of n-alkanes.
\newblock {\em J. Comput.\ Phys.}, 23:327 -- 341, 1977.

\bibitem{ScEK94}
J.~Schlitter, M.~Engels, and P.~Kr\"uger.
\newblock Targeted molecular dynamics: {A} new approach for searching pathways
  of conformational transitions.
\newblock {\em J. Mol.\ Graphics}, 12:84--89, 1994.

\bibitem{SiPa05}
N.~Singhal and V.~S. Pande.
\newblock Error analysis and efficient sampling in {M}arkovian state models for
  molecular dynamics.
\newblock {\em J. Chem.\ Phys.}, 123:204909, 2005.

\bibitem{Vand06}
E.~Vanden-Eijnden.
\newblock Transition path theory.
\newblock In K.~Binder, G.~Ciccotti, and M.~Ferrario, editors, {\em Computer
  Simulations in Condensed Matter: From Materials to Chemical Biology,
  Volume~2}, Lecture Notes in Physics, pages 453--493. Springer, Berlin, 2006.

\bibitem{Vand08}
E.~Vanden-Eijnden.
\newblock Theoretical equivalence between original string and swarm string in
  the short lag limit.
\newblock Unpublished work, 2008.

\bibitem{VaHe08}
E.~Vanden-Eijnden and M.~Heymann.
\newblock The geometric minimum action method for computing minimum energy
  paths.
\newblock {\em J. Chem.\ Phys.}, 128:061103, 2008.

\bibitem{VaVe09}
E.~Vanden-Eijnden and M.~Venturoli.
\newblock Revisiting the finite temperature string method for the calculation
  of reaction tubes and free energies.
\newblock {\em J. Chem.\ Phys.}, 130:194103, 2009.

\bibitem{VVCE08}
E.~Vanden-Eijnden, M.~Venturoli, G.~Ciccotti, and R.~Elber.
\newblock On the assumptions underlying milestoning.
\newblock {\em J. Chem.\ Phys.}, 129:174102, 2008.

\bibitem{YaBR09}
S.~Yang, N.~K. Banavali, and B.~Roux.
\newblock Mapping the conformational transition in {S}rc activation by
  cumulating the information from multiple molecular dynamics trajectories.
\newblock {\em PNAS}, 106:3776--3781, 2009.

\bibitem{ZhYG09}
J.~Zhang, P.~L. Yang, and N.~S. Gray.
\newblock Targeting cancer with small molecule kinase inhibitors.
\newblock {\em Nat.\ Rev.\ Cancer}, 9:28--39, 2009.

\bibitem{Zhao09}
R.~Zhao.
\newblock Mftp code, 2009.
\newblock URL {\tt http://bionum.cs.purdue.edu/mftp}.

\end{thebibliography}
\bibliographystyle{abbrv}


\section*{Captions}
Figure~1: A schematic illustration of a poor choice of collective variables.
The horizontal axis is collective variables, and the vertical axis is
unrepresented degrees of freedom.
The collective variables fail to indicate the progress of the reaction.

Figure~2: Shading indicates contours of free energy, thin curves denote
isocommittors, ellipses
enclose concentrations of crossing points from reactive trajectories, and
the thick curve is the center.

Figure~3: Minimum energy path obtained using the simplified string method.
The initial path is the straight line between $(-1,0)$ and $(1,0)$.
The path is discretized into $J+1$ images.
Four figures are generated using $J+1=10$, $20$, $40$, $80$ images,
respectively.

Figure~4: Maximum flux transition path obtained using the semi-implicit
simplified string method.
Here we used the same initial path and the same stopping criterion
for convergence as for \ref{fig:mep}.
The MFTPs are generated using 20 images at 3\,K, 30\,K, 300\,K, 3000\,K,
30000\,K, respectively (which roughly correspond to $\beta^{-1} = 0.006, 0.06, 0.6, 6, 60$ kcal/mol.)
The contour lines are separated by 0.25\,kcal/mol.

Figure~5: Maximum flux transition path and minimum free energy
path from $C_{7\mathrm{eq}}$ to $C_{7\mathrm{ax}}$ for alanine dipeptide
in vacuum at 300\,K.
Triangles are images of the initial path;
rectangles are the images of the maximum flux transition path;
and circles are the images of the minimum free energy path.
The contours are those for the zero-temperature free energy (adiabatic energy).
The contour lines are separated by 0.6\,kcal/mol.

Figure~6: Maximum flux transition path and minimum free energy
path for alanine dipeptide from $C_{\mathrm{7eq}}$ to $C'_{\mathrm{7eq}}$
passing by $C_{\mathrm{7ax}}$ in vacuum at 300\,K.
The figure is generated using 40 images.
Triangles are the images for the initial path;
rectangles are the images of the maximum flux transition path;
and circles are the images of the minimum free energy path.
The contours are those for the zero-temperature free energy.
The contour lines are separated by 0.6\,kcal/mol.

Figure~7: Maximum flux transition paths for alanine dipeptide in solution.
The transition paths are calculated by the semi-implicit simplified string method
with the nearby straight lines as initial paths.
\end{document}